\documentclass[preprint,a4paper,pra,showpacs]{revtex4-1}
\pdfoutput=1
\usepackage{amssymb}
\usepackage{amsfonts,amsbsy}
\usepackage{graphicx}

\def\e{\mathrm{e}}

\def\d{\mathrm{d}}

\def\beq{\begin{equation}}
\def\eeq{\end{equation}}

\def\vx{\mathbf{x}}

\def\E{\mathsf{E}\,}

\def\bx{\boldsymbol{x}}
\def\by{\boldsymbol{y}}

\def\bv{\boldsymbol{v}}

\newcommand{\bn}{\begin{eqnarray}}
\newcommand{\en}{\end{eqnarray}}
\newcommand{\gbar}{\bar \gamma}

\newcommand{\ttexp}[1]{10^{#1}}  

\newcommand{\eqn}[1]{(\ref{eqn:#1})}
\newcommand{\lab}[1]{\label{eqn:#1}}
\newcommand{\inter}[1]{\quad \textrm{#1} \quad}

\def\XXint#1#2#3{{\setbox0=\hbox{$#1{#2#3}{\int}$}
\vcenter{\hbox{$#2#3$}}\kern-.5\wd0}}

\begin{document}

\title{Scalar decay in a three-dimensional chaotic flow}
\author{K. Ngan}
\email{keith.ngan@metoffice.gov.uk}
\affiliation{Met Office, Exeter EX1 3PB, UK}
\author{J. Vanneste}
\email{J.Vanneste@ed.ac.uk}
\affiliation{School of Mathematics and Maxwell Institute for Mathematical Sciences, University of Edinburgh, Edinburgh EH9 3JZ, UK}
\date{March 2011}

\begin{abstract}
  The decay of a passive scalar in a three-dimensional chaotic flow is
  studied using high-resolution numerical simulations. The (volume-preserving) flow considered is a three-dimensional extension of the randomised alternating sine flow employed extensively in studies of mixing in two dimensions. It is used to show that theoretical predictions for two-dimensional flows with
  small diffusivity carry over to three dimensions even though the
  stretching properties differ significantly.  The variance decay
  rate, scalar field structure, and time evolution of statistical
  moments confirm that there are two distinct regimes of scalar decay:
  a locally controlled regime, which applies when the domain size is comparable to
  the characteristic lengthscale of the velocity field,
  and a globally controlled regime, which when applies when the domain is larger.  
  Asymptotic predictions for the
  variance decay rate in both regimes show excellent agreement with
  the numerical results. Consideration of both the forward flow and its time reverse makes it possible to compare the scalar evolution in flows with one or two expanding directions; simulations confirm the theoretical prediction that the decay rate of the scalar is the same in both flows, despite the very different scalar field structures.

\end{abstract} 


\maketitle

\section{Introduction}

In studies of fluid mixing, much attention has been devoted to chaotic-advection flows, characterised by relatively simple (smooth and
divergence-free) velocity fields but complex, chaotic particle
trajectories. Flows of this type appear in several contexts: Stokes
flows \citep{otti89}, turbulent flows with large Schmidt number (in the so-called Batchelor regime \citep{batc59}), geophysical flows \citep{hayn-angl,ngan-shep98a}, and elastic turbulence \citep{burg-et-al}. 
One aspect is of particular interest: the decay, through the combined
effect of advection and diffusion, of the concentration of passive
scalars released in such flows. In bounded domains this decay is
exponentially fast; one of the main problems is then to relate the
corresponding decay rate to the flow characteristics and to the
diffusivity.

For small diffusivity $\kappa \to 0$, a clear distinction emerges
between flows that are everywhere chaotic (e.g.\ uniformly
hyperbolic), and flows with regular regions (e.g. containing elliptic
islands \citep{piko-popo} or incorporating no-slip boundary
conditions \citep{lebe-turi,salm-hayn}): the decay rate of passive
scalars tends to a non-zero  limit as $\kappa \to 0$ in the first
case, whereas it tends to zero in the second case. In this paper, we
concentrate on the first case and, more specifically, on stochastic
models of chaotic flows. These models, in which the complex time
dependence of realistic flows is represented by random
processes \citep{krai68,krai71}, have been the subject of sustained
research from the mid 1990s onward \citep{falk-et-al}. Remarkably, several
theoretical results are now available which relate properties of the
scalar decay, notably the decay rate, to flow properties such as
stretching
statistics \citep{anto-et-al,balk-foux,son99,fere-et-al,fere-hayn,sche-et-al04,tsan-et-al05a,hayn-v05}.
These results, which we review in Sec.~\ref{sec:decay} below, have
been confirmed by numerical simulations \citep{tsan-et-al05a,hayn-v05}.
Confirmation, however, has been limited to two-dimensional flows. The main aim of the present paper is to test the theoretical results against numerical simulations of three-dimensional flows. 

The passage from two to three dimensions adds significantly complexity
to the problem. Numerically, the high resolutions needed to capture
the fine scalar scales generated for small diffusivities make the 3D
case much more computationally demanding than 2D. Physically, the
dynamics of scalars in 3D flows are also much richer than in 2D flows:
incompressibility provides a strong constraint in 2D, leading in
particular to equal and opposite (stretching) Lyapunov exponents; in
3D incompressibility represents a looser constraint, imposing only
that the sum of the three Lyapunov exponents vanish. The existence of three distinct Lyapunov exponents also has a profound influence on  the concentration field. Depending on whether the intermediate Lyapunov exponent is positive or negative, the concentration field is dominated either by quasi-two-dimensional structures (``pancakes'') or by quasi-one-dimensional structures (``needles''). Interestingly, the most recent theoretical results about the scalar decay rate \citep{tsan-et-al05a,hayn-v05} are insensitive to this difference in structure and make predictions that are independent of the sign of the intermediate Lyapunov exponent. Our simulations confirm the validity of this conclusion.

A well-known difference between mixing in 2D and 3D is that chaotic particle trajectories are only possible in 2D flows if they are time dependent, whereas they exist in time-independent 3D flows. Time-independent 3D flows, however, are similar to one-degree-of-freedom Hamiltonian maps \citep{mack94}, and do not  lead to  global chaos but rather to a mixture of chaotic and (possibly small) regular regions, with the latter controlling scalar decay in the long-time limit. The flows with random time dependence that we consider do not have this type of intricate structure: they are statistically homogeneous and globally mixing. This is one of the aspects which differentiate this paper form earlier work on mixing in 3D, in particular the numerical simulations of time-independent flows in 
\citep{tous-et-al95} and \citep{tous-et-al00}.

Qualitatively, the mixing properties of sufficiently chaotic flows are independent
of the flow details. It is therefore advantageous to devise a
random-flow model that is easy to implement and analyse. In 2D, this
role is played by the randomised sine map \citep{pier94,pier00} which
has been used extensively in studies of passive and reactive scalars.
In this paper, we propose a straightforward 3D extension of the sine
map. With the parameters chosen, this map has two positive and one
negative Lyapunov exponent, which yield concentration fields with
pancake-like structure. We use the associated inverse map as a model for flows
with two negative and one positive Lyapunov exponent which generate
needle-like concentration fields. The extension to 3D is
not uniquely defined. Other versions of the 3D map are less mixing in
that they do not stretch certain fixed directions. We briefly comment
on the impact this property has on the scalar decay.

The plan of the paper is as follows. In Sec.~\ref{sec:decay}, we
review the theoretical predictions for the decay rate of the scalar
concentration. Following \citep{hayn-v05}, we emphasise the
existence of two regimes of scalar decay: a locally controlled regime
relevant to flows whose typical scale is comparable to the size of the
(periodic) domain considered, and a globally controlled regime
relevant to flows of smaller scale (see also
\citep{tsan-et-al05a}). The three-dimensional sine map is
introduced in Sec.~\ref{sec:formulation}. There we examine the
stretching properties of the map and describe the numerical procedure
employed for simulations of the scalar evolution.
Secs.~\ref{sec:local-control} and \ref{sec:trans-glob-contr} present
results of simulations carried out in different domain sizes. This
makes it possible to explore both locally and globally controlled
regimes and contrast differences in the decay rate as well as in the
structure of the concentration field.  We also compare concentration
fields obtained with the map and its inverse. The paper concludes with
a discussion in Sec.~\ref{sec:discussion}.

\section{Decay rate} \label{sec:decay}

The evolution of the concentration $C(\bx,t)$ of a passive scalar
released in a flow $\bv(\bx,t)$ is governed by the
advection--diffusion equation 
\beq \lab{ad}
\partial_t C + \bv \cdot \nabla C = \kappa \nabla^2 C, \quad 
\eeq
with initial conditions $C(\bx,0)=C_0(\bx)$.  When non-dimensionalised
using characteristic length and velocity scales, this problem retains
its form, with $\kappa$ now representing the inverse P\'eclet number
rather than the diffusivity. This non-dimensional interpretation will
be used in what follows.

We focus on incompressible flows, $\nabla \cdot \bv
=0$, generated by specific random processes. These flows are assumed to have homogeneous and stationary statistics and to be
spatially smooth, that is,  $\|\bv(\bx)-\bv(\bx')\| \propto \|\bx-\bx'\|$ as $\|\bx-\bx'\| \to 0$. Periodic boundary conditions
are applied to \eqn{ad}. Since \eqn{ad} is left unchanged when a constant is added to $C$,  there is no loss of generality in assuming that the concentration field averages to zero:
\beq 
\int C(\bx,t) \, \d \bx = \int C_0(\vx) \, \d \bx = 0.
\eeq

By analogy with the finite-dimensional case the concentration $C(\bx,t)$ can be expected to decay exponentially in the long-time limit:
\beq \lab{adle} 
C(\bx,t) \sim D(\bx,t) \e^{-\lambda t} \ \ \ \textrm{as} \ \ \to \infty
\eeq 
for almost all realisations of $\bv(\bx,t)$, Here the deterministic decay rate $\lambda$ is best interpreted as (minus) the Lyapunov exponent of the linear system \eqn{ad}, and $D(\bx,t)$ is a stationary function termed strange eigenmode  \citep{pier94} (see \citep{hall-yuan} for rigorous results). We emphasise that $\lambda$ is the Lyapunov exponent of the infinite-dimensional random system \eqn{ad} and should not be confused with the Lyapunov exponent of the three-dimensional linear system governing the separation of nearby particles in the velocity field $\bv$. We also note that $\lambda^{-1}$ is not the only useful time scale characterising the scalar decay: a dissipation time scale can for instance be defined as the time for some norm of $C$ to be reduced from its initial value by a given fraction \citep{fann-et-al04}; this, however, characterises the early-time behaviour rather than the long-time behaviour on which we focus. 

A striking feature of the decay rate is that it tends to a non-zero
value in the limit of zero diffusivity (infinite P\'eclet number), $\lim_{\kappa \to 0} \lambda =
\lambda_0 \not= 0$, provided the flow is sufficiently mixing. This
raises the question of the relationship between the asymptotic decay
rate, $\lambda_0$, and the statistical properties of the flow,
$\bv(\bx,t)$. 

In a series of papers \citep{anto-et-al,balk-foux,falk-et-al,son99},
$\lambda_0$ has been related to the large-deviation statistics of the
finite-time Lyapunov exponents of $\bv(\bx,t)$. These are defined as
\beq \lab{ftle}
h = t^{-1} \log \| \by \|,
\eeq
where $\by$ denotes the separation between nearby trajectories, and are distributed according to a pdf, $p(h;t)$ say. $\lambda_0$ may be expressed in terms of the rate function 
\beq \lab{rate}
g(h) = -\lim_{t \to \infty} \log p(h;t)
\eeq
associated with this pdf.
Alternatively, it may be expressed in terms of the generalised Lyapunov
exponent
\beq \lab{ell}
\ell(q)=\lim_{t \to \infty} t^{-1} \log \E \| \by \|^q,
\eeq
where $\E$ denotes ensemble average. These formulations are equivalent since
$g(q)$ and $\ell(q)$ are related by a Legendre transform \citep{ott02,v10}:
\beq \lab{legendre}
\ell(q) = \sup_h \left( q h - g(h) \right).
\eeq

Results of this type are not general, however. In particular, they do not apply to flows whose typical scale is much
smaller than the domain size \citep{fere-et-al,fere-hayn,sukh-pier02a,thif-chil}. In fact there are two different regimes of scalar decay \citep{sche-et-al04,tsan-et-al05a,hayn-v05}. Flows whose spatial scale is comparable to the domain size are in a \textit{locally controlled} regime, in which $\lambda_0$ depends only on the stretching statistics of $\bv(\bx,t)$ and is given in terms of $g(h)$ or $\ell(q)$. Flows with smaller scale, on the other hand, are in a \textit{globally controlled} regime, in which $\lambda_0$ depends on global flow features including the domain size. 

In  \citep{hayn-v05}, the two regimes are related to two
different parts of the spectrum of the (deterministic) linear operator
governing the evolution of the covariance $\E C(\bx,t) C(\bx',t)$ in
Kraichan--Kazantsev and renewing flows \footnote{``Kraichan--Kazantsev
  flows'' are obtained in the limit of zero correlation time for the
  velocity field (i.e.\ a white-in-time velocity field);  ``renewing flows'' or  ``renovating flows'' are described by
  independent, identically distributed random processes that become
  completely decorrelated after a finite time interval.}.
The smallest non-zero
eigenvalue determines $\gbar$, the decay rate of the ensemble-averaged variance:
\beq
\bar \gamma = - \lim_{t \to \infty} t^{-1} \log \, \E \int C^2(\bx,t) \, \d \bx.
\eeq
In the locally controlled regime, $\bar
\gamma$ belongs to the continuous spectrum of the operator obtained for $\kappa=0$, while in the globally
controlled regime, $\bar \gamma$ represents a discrete eigenvalue.

Because of this difference in the eigenvalues, the decay rates are qualitatively different in the two regimes.  In the locally controlled regime
the asymptotic form of $\bar \gamma$ for $\kappa \to 0$ is
\beq \lab{localdecay}
\bar \gamma_\mathrm{local} = -\ell(q_*) +\frac{2 \pi^2 \ell''(q_*)}{\log^2 \kappa}  + o(1/\log^2 \kappa),
\eeq 
where $q_*$ is the minimum of $\ell(q)$ [so that $\ell(q_*)=-g(0)$, see \eqn{legendre}]. In the
globally controlled regime, no such simple expression exists, and the
leading-order term can be obtained only by solving the eigenvalue
problem for the covariance. However, in the limit of flow scale much
smaller than domain size, a useful estimate is provided by
homogenisation theory, which approximates \eqn{ad} by a diffusion
equation with an effective diffusivity
$\kappa_{\mathrm{eff}}$ \citep{majd-kram}, whence
\beq \lab{globaldecay}
\bar \gamma_\mathrm{global} \sim \bar \gamma_\mathrm{hom} = \frac{8 \pi^2 \kappa_\mathrm{eff}}{L^2},
\eeq 
where  $L$ is the largest of the domain periods.

Note that \eqn{localdecay} and \eqn{globaldecay} describe the decay rate of the
ensemble-averaged variance.  The decay rate of a single
realisation is given by
\beq \lab{gamma}
\gamma = - \lim_{t \to \infty} t^{-1}  \log  \int C^2(\bx,t) \, \d \bx = 2 \lambda
 \eeq 
on using \eqn{adle}.
 In general, $\gamma \not= \bar \gamma$, the difference marking the
 intermittency of the scalar decay \citep{v06a}. For the flows considered here,
 differences between $\gamma$ and $\bar \gamma$ are negligible (see
 Sec.~\ref{sec:local-control}). Indeed we will confirm that \eqn{localdecay}
 can be used to predict the variance decay in single realisations.  (The
 equality of $\gamma$ and $\bar \gamma$ in the homogenisation limit
 \eqn{globaldecay} is clear since the spatially homogenised equation is
 deterministic.)

 The theory summarised above holds in any dimension \citep{hayn-v05}.
 Until now, numerical verification (of \eqn{localdecay} and
 \eqn{globaldecay} in particular) has been restricted to two
 dimensions. In this paper we assess the applicability to
 three-dimensional flows.

 The additional dimension introduces interesting complexities.  In two dimensions, incompressibility strongly constrains the distribution of finite-time Lyapunov exponents, leading to the symmetry property $\ell(q)=\ell(-q-2)$. As a result,  $q_*=-1$ and the leading-order of \eqn{localdecay}, which in general reads  $\bar \gamma \sim - \ell(q_*) = g(0)$, can be alternatively computed as $\bar \gamma \sim - \ell(-1)$ (it is in this form that the decay rate appears in some of the theories \citep{anto-et-al,balk-foux,tsan-et-al05a}). In dimensions $d\ge2$, by contrast, there is no such symmetry property; rather what hold are the relationships \citep{v10}
%
\beq \lab{ellsym}
\ell^-(q) = \ell(-q-d), \inter{and hence} g^{-}(h)=g(-h)-d h
\eeq
between the stretching statistics of a flow and those of the
time-reversed flow (denoted by~$^-$). A consequence is that
\eqn{localdecay} predicts that the decay rate of a scalar is the same
in a flow and in the time-reversed flow. This is a remarkable
prediction for $d = 3$ because  many features of the stretching in a
flow and its time-reverse are different. In particular, the intermediate Lyapunov exponent changes sign; thus, if a flow has
one contracting and two expanding directions, the time-reverse has one
expanding and two contracting directions. As a result, the scalar fields have very different structures in the two flows, with typical concentration isosurfaces taking the shape of (nearly two-dimensional) pancakes in one case  and the shape of (nearly one-dimensional) needles in the other. Nonetheless, the scalar concentration  decays at the same rate, as we verify below
\footnote{Note that \citep{balk-foux} predicts a decay rate for $d=3$ which differs from $\gamma \sim - \ell(q_*) = g(0)$ and depends on whether a flow has one or two expanding directions.}.

\section{Three-dimensional sine map} \label{sec:formulation}

\subsection{Formulation}

The randomised two-dimensional sine map \citep{pier94,pier00} has become a standard tool for the study of mixing by spatially smooth flows.
It is given by
\beq \lab{sine2d}
x_{n+1} = x_n + a \sin (y_n + \phi_n), \quad y_{n+1} = y_n + b \sin(x_{n+1} + \psi_n),
\eeq
where $a$ and $b$ are constant parameters and $\phi_n$ and $\psi_n$
are independent random angles uniformly distributed in $[0,2\pi]$. The domain considered is periodic so the right-hand sides in \eqn{sine2d} are taken modulo $2 \pi$. 
This map provides the positions at times $n \tau, \,
n=1,2\cdots$ of particles advected by a succession of shear flows alternating in direction. Its application to \eqn{ad} is straightforward: advection of a scalar field by \eqn{sine2d} can be implemented very efficiently using a lattice
representation; diffusion for small $\kappa$ can be incorporated as
smoothing by the heat kernel $\propto \exp(-(x^2+y^2)/(4 \kappa
\tau))$ (or rather the periodised version thereof).

In what follows, we use a straightforward three-dimensional generalisation
of \eqn{sine2d}, namely
\beq \lab{sine3d}
x_{n+1} = x_n + a \sin(y_n + \phi_n), \quad 
y_{n+1} = y_n + b \sin(z_n + \psi_n), \quad
z_{n+1} = z_n + c \sin(x_{n+1} + \varphi_n),
\eeq
where $a, \, b$ and $c$ are constant, and $\phi_n, \, \psi_n$ and
$\varphi_n$ are independent uniformly distributed random phases in
$[0,2\pi]$. This map, which corresponds to successive applications of
shear flows in the $x$-, $y$- and $z$-directions, preserves volume
and has homogeneous statistics. As in two dimensions, these shear flows are simple enough that they can be integrated explicitly to yield \eqn{sine3d}.
The stretching statistics of the map are
controlled by the Jacobian matrix, given at the origin by
\beq \lab{sine3dA}
A_n = \left( \begin{array}{ccc}
1 & a \cos \phi_n & 0 \\
0 & 1 & b \cos \psi_n \\
c \cos (a \sin\phi_n +\varphi_n) & a c  \cos \phi_n  \cos (a \sin\phi_n +\varphi_n) 
& 1 \end{array} \right).
\eeq 
The three Lyapunov exponents are computed numerically from the singular values of $A_n A_{n-1} \cdots A_0$. In what follows, we fix $a=b=c=\pi$. In this case, the  Lyapunov exponents are found to be 1.04, 0.58 and
$-1.62$. Thus the map \eqn{sine3d} has one contracting and two expanding
directions. 

The inverse map,
\beq \lab{sine3dinv}
z_{n+1}=z_n+c \sin(x_n + \varphi_n), \quad
y_{n+1}=y_n+b \sin(z_{n+1} + \psi_n), \quad
x_{n+1}=x_n + a \sin(y_{n+1} + \phi_n),
\eeq 
corresponds to the time-reversed flow.  It has Jacobian matrix $A_n^{-1}$, and for $a=b=c=\pi$ Lyapunov exponents $1.62$, $-0.58$ and $-1.04$,
 hence one expanding and two contracting directions. 

Three-dimensional generalisations of \eqn{sine2d} are not
uniquely determined. Maps differing from \eqn{sine3d} and
\eqn{sine3dinv} via permutations of the variables can be constructed.
However, they may not be transitive, in the sense that their Jacobian
matrices have invariant directions independent of the random phases.
Such maps may not be sufficiently mixing for predictions such as
\eqn{localdecay} to hold, even if the largest Lyapunov exponent is
positive. We return to this point in Sec.~\ref{sec:discussion}.

\subsection{Theoretical predictions}
\label{sec:theor-pred}

We now consider the predictions \eqn{localdecay}--\eqn{globaldecay}
for the three-dimensional map \eqn{sine3d} and its inverse
\eqn{sine3dinv}. We choose the parameters $a=b=c=\pi$ and take
$\tau=1$.

\begin{figure}
\begin{center}
\includegraphics[height=6cm]{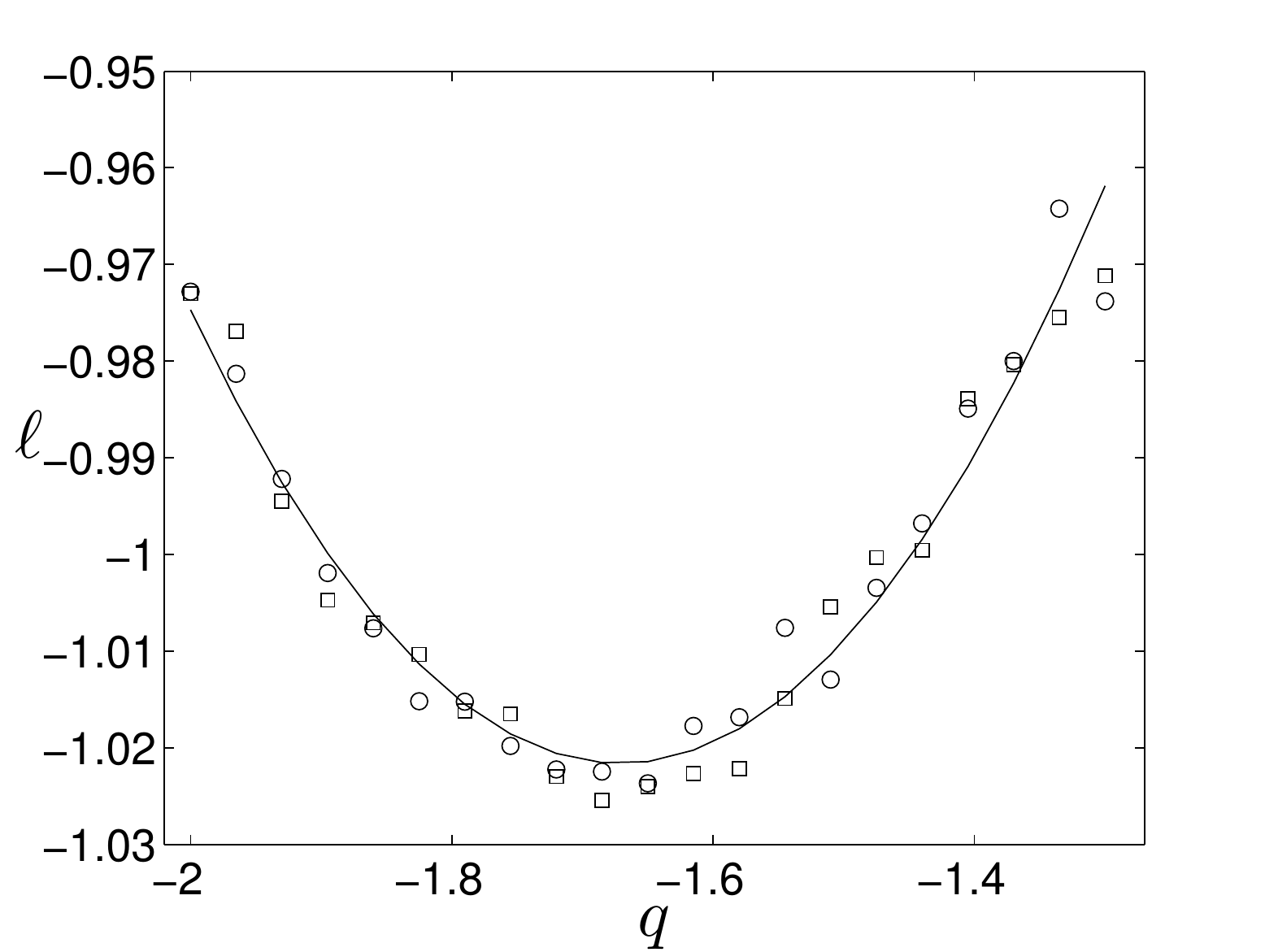}
\caption{Generalised Lyapunov exponents for the map \eqn{sine3d} with
  $a=b=c=\pi$. The results of two Monte Carlo computations (circles
  and squares) are shown together with the best fit by a parabola.} \label{fig:ellqpi}
\end{center}
\end{figure}

In order to find the coefficients in the asymptotic formula
\eqn{localdecay}, we use a Monte Carlo approach to compute $\ell(q)$
numerically in the vicinity of its minimum, $q_*$. Because the
expectation, $\E \| \by \|^q$, is dominated by rare realisations, the
sampling required is rather delicate; we therefore use importance
sampling, namely the random resampling method described in 
\citep{v10}.  In particular, we use an ensemble of $5 \times 10^4$
realisations to estimate $\ell(q)$ over a range bracketing $q_*$, $q
\in [-2,-1.4]$; $\ell(q_*)$ and $\ell''(q_*)$ follow from a parabolic
fit. Some numerical results are shown in Figure \ref{fig:ellqpi}.
Rough error bars are obtained by repeating this procedure 10
times. This provides the estimates $ \ell(q_*)=1.022 \pm 0.002$ and
$\ell''(q_*)=0.93 \pm 0.09$. Introducing these values
into \eqn{localdecay} then yields the estimate
%
\beq \lab{localdecaypi}
\bar \gamma_\mathrm{local} = 1.022 (\pm 0.002)  + \frac{18.36 (\pm 1.78)}{\log^2 \kappa}.
\eeq
for the variance decay rate in the locally controlled regime. This
is tested in Sec.~\ref{sec:local-control}.


The corresponding prediction for the inverse map \eqn{sine3dinv} is exactly the
same as that for the direct map \eqn{sine3d}.  Indeed, Eq.~\eqn{ellsym} implies
that $\ell^-(q)$ is the reflection of $\ell(q)$ about the line $q=-3/2$.  Therefore, $\ell^-(q_*)=\ell(q_*)$, $(\ell^{-})''(q_*)=\ell''(q_*)$ and hence $\gamma_\mathrm{local}$ is unchanged.

To examine both locally and globally controlled regimes of scalar
decay, we consider the scalar evolution for the forward map in a
periodic cube of (total) length $2 \pi P$ with
 $P=1,2\cdots$; that is, the flow has unit cells of length $2\pi$ but the right-hand sides of \eqn{sine3d} are taken modulo $2\pi P$. The asymptotic result \eqn{localdecay} is expected to hold for small values of $P$, while the homogenisation approximation \eqn{globaldecay} should be valid for $P \gg 1$. 
The effective
diffusivity $\kappa_\mathrm{eff}$ appearing in \eqn{globaldecay}, is easily found by recalling that
the variance of one of the coordinates, $x_n$ say, satisfies $\E x_n^2
= 2 \kappa_\mathrm{eff} n$; computing the variance from \eqn{sine3d}
gives $\kappa_\mathrm{eff}=a^2/4$. Hence
\beq \lab{globaldecaypi}
\bar \gamma_\mathrm{global} \sim \bar \gamma_\mathrm{hom} = \frac{\pi^2}{2 P^2}.
\eeq
for large $P$. 

We wish to estimate the value of $P$ at which the transition from local to global control occurs. The predicted decay rate $\gbar$ is determined by whichever of the local and global decay rates is smaller.  %
Comparing \eqn{localdecaypi} with \eqn{globaldecaypi}  we therefore expect local control for $P=1,2$ and global control for $P \ge 3$.

\subsection{Numerical procedure}

Numerical simulations are performed on a regular grid of $N^3$ points.
The advection uses the 3-D sine map \eqn{sine3d}, while for
numerical efficiency the diffusion is applied in spectral space using
a Fast Fourier Transform. We have confirmed that an explicit
finite-difference step yields essentially indistinguishable results.
The code has been tested by reproducing results of \citep{hayn-v05} for the 2-D sine map \eqn{sine2d}.

Two sets of initial conditions are considered. In
most simulations we use
\beq
C(\bx,0) = \sin(x/P) \sin(y/P) \sin(z/P). \lab{ics_local} 
\eeq
In Sec.~\ref{sec:trans-glob-contr}, we also use the initial condition
\beq
C(\bx,0) = \sin(z/P) \lab{ics_global}
\eeq
corresponding to one of the gravest modes of the diffusion operator in the domain.

The simulations need to resolve spatial scales in the range between
the box size, $L_0$, and the tracer microscale, $L_\kappa$. The latter
scale is estimated by matching the diffusive and stretching
timescales, yielding $L_\kappa = \sqrt{\kappa/\lambda}$, where
$\lambda$ is the largest Lyapunov exponent of $\bv$. Thus we require
$L_\kappa \gtrsim \Delta x$, where $\Delta x=L_0/N$ is the grid spacing, or
\beq \kappa \gtrsim (\frac{2\pi}{N})^2\lambda.  \eeq
In the simulations of Secs.~\ref{sec:local-control} and
\ref{sec:trans-glob-contr}, $N=1024$. This implies a critical value
of the diffusivity, $\kappa_c = O(\ttexp{-5})$.

For $P>1$, the numerical simulations are performed on the $N^3$ grid
but with $P^3$ replicas of the velocity field.  Taking $P>1$ amounts
to a redefinition of $L_0$. For simplicity, however, it is convenient
to keep $L_0=2\pi$ fixed and rescale the equations.
 This allows us to reuse the code described above with minimal changes. More
precisely, we use the modified map
\bn 
&x_{n+1} = x_n + \displaystyle{\frac{a}{P}} \sin(P y_n + \phi_n), \quad 
y_{n+1} = y_n + \displaystyle{\frac{b}{P}} \sin(P z_n + \psi_n), & \nonumber  \\
&z_{n+1} = z_n + \displaystyle{\frac{c}{P}} \sin(P x_{n+1} + \varphi_n),& 
\lab{fwd_modified}
\en
with the scaling $ \kappa \mapsto P^2\kappa$ and the right-hand sides taken modulo $2 \pi$.
Note that the Jacobian is unchanged.

\section{Local control: $P = 1$}
\label{sec:local-control}

 \begin{figure}[t]
 \begin{center}
 \begin{tabular}{cc}
 (a) & (b) \\
\includegraphics[scale=0.42]{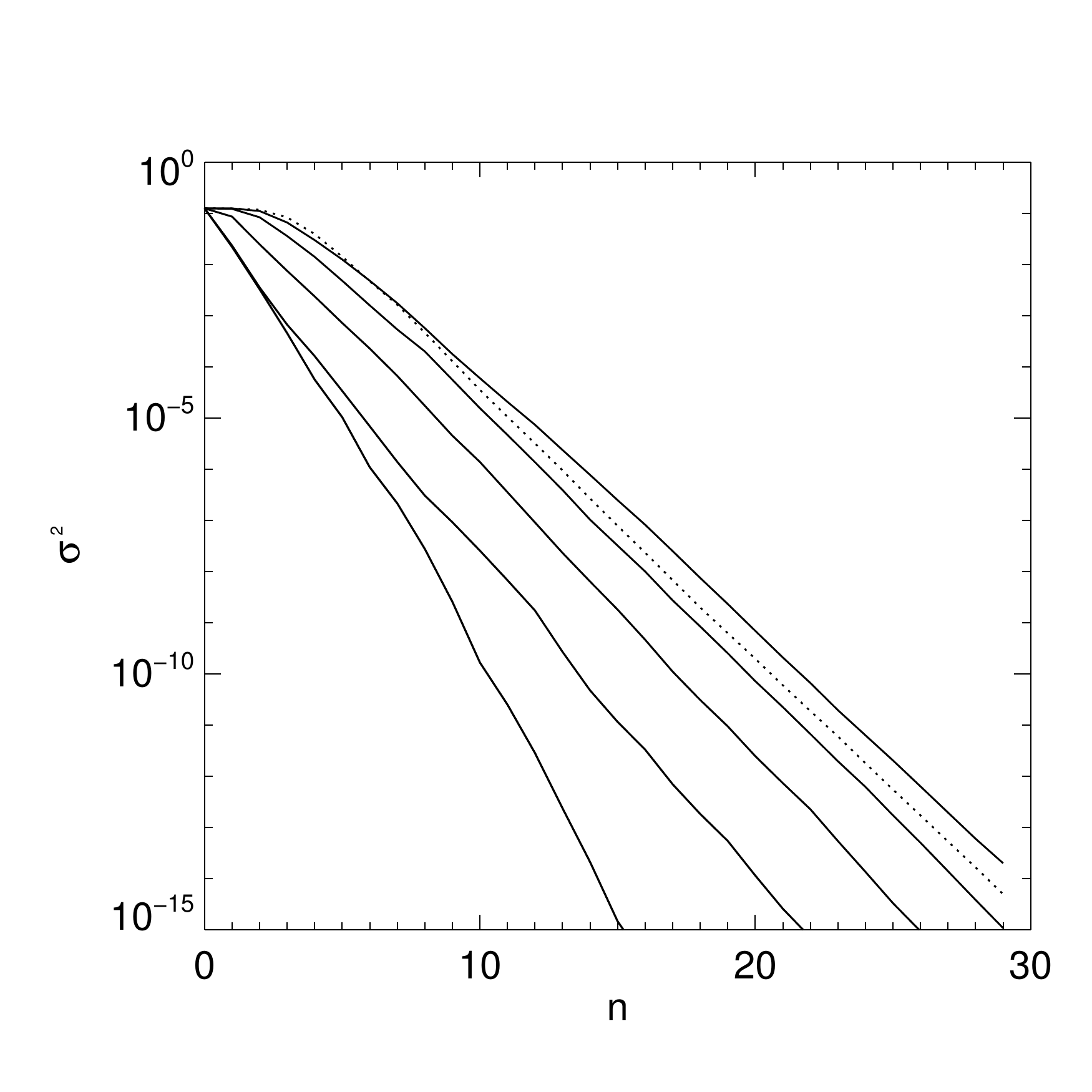} &
\includegraphics[scale=0.42]{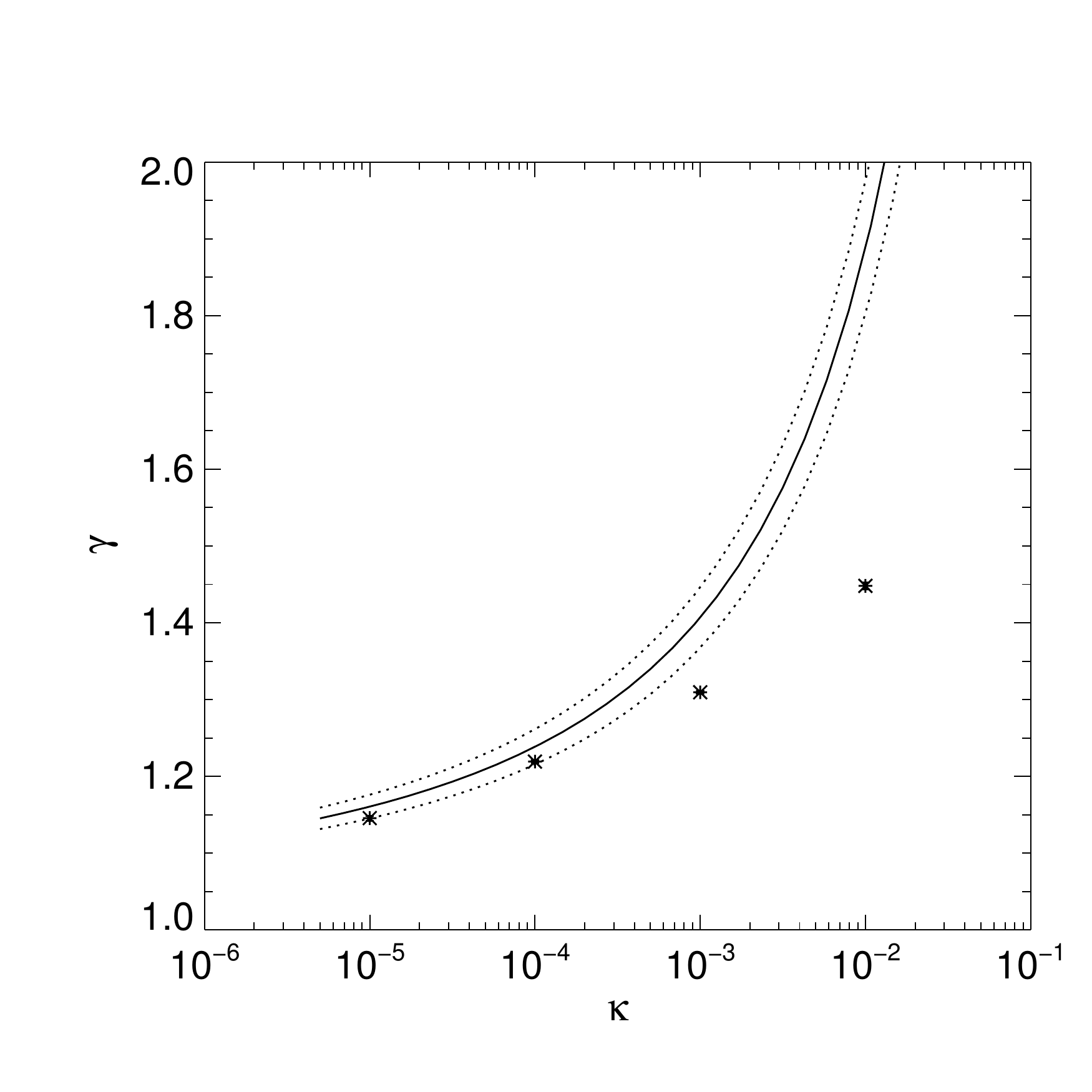}
\end{tabular}
\caption{Variance decay for $P=1$. (a): variance $\sigma^2$ versus $n$ for the
  forward map  with $\kappa=10^{-5},\, 10^{-4}, \,
  10^{-3}, \, 10^{-2}, \, 10^{-1}$ (solid lines, from top to bottom),
  and for the inverse map with $\kappa=10^{-4}$ (dashed line). (b):
  decay rate $\gamma$ versus $\kappa$ for the forward map. The decay
  rates, obtained from a least-squares fit over $n\in[5,30]$ (symbols, with error bars corresponding to the standard error), are compared with the prediction $\gbar_\mathrm{local}$ in  \eqn{localdecaypi} (solid line, with dashed lines indicating the error estimates.)} 
\label{fig:vardecay}
\end{center}

\end{figure}

In this section, we focus on the locally controlled regime and take $P=1$.
We report on simulations of the forward map \eqn{sine3d} for several values of $\kappa$ in the range $10^{-5}$--$10^{-1}$, and of the inverse map \eqn{sine3dinv} for $\kappa=10^{-4}$.
%

The evolution of the variance
\begin{equation}
  \lab{vardef}
  \sigma^2(n)=\int C^2(\bx,n) \, \d\bx
\end{equation}
in these simulations is shown in Fig.~\ref{fig:vardecay}(a). 
As with the two-dimensional sine map \eqn{sine2d}, the variance decays exponentially after a brief initial transient, in agreement with the strange-eigenmode prediction \eqn{adle}. These results are insensitive to resolution: simulations at the lower resolution $N=512$ (not shown) are almost identical. Also, there is little variability between different realisations.  
For an
ensemble of 8 realisations evaluated at
$N=512$ (not shown), the decay rate of the variance for $\kappa=\ttexp{-4}$ varies by about $2\%$. This difference is of the same order of
magnitude as the difference between the decay rates for $N=512$ and
$N=1024$.  We therefore conclude that the intermittency is negligible
and henceforth restrict attention to the single-realisation decay rate $\gamma$ which we identify with $\bar \gamma$.

The forward and inverse maps lead to almost identical decay rates.
Specifically, for $\kappa=10^{-4}$, the decay rates for the forward
and inverse maps are 1.22 and 1.21, respectively. This supports the
claim made in Sec.~\ref{sec:theor-pred} that their decay rates are
equal despite differences in their stretching properties.

The dependence of $\gamma$ on $\kappa$ for
the forward map is shown in Fig.~\ref{fig:vardecay}(b). For each value of $\kappa$, $\gamma$ is
obtained from a least-squares fit over the time interval, $n \in [5,30]$.
There is excellent agreement with the asymptotic estimate \eqn{localdecaypi} for $\gbar_\mathrm{local}$. This confirms the applicability of the asymptotic result \eqn{localdecay} to the locally controlled regime in more than two dimensions. The agreement deteriorates rather rapidly for larger $\kappa$, but  this is to be expected since the theory assumes $\kappa\to0$ and neglects terms that are $o(1/\log^2 \kappa)$.


 \begin{figure}[t]
 \begin{minipage}{5.175cm}
(a) 
\includegraphics[scale=0.32,clip=true,trim=50 50 50 50]{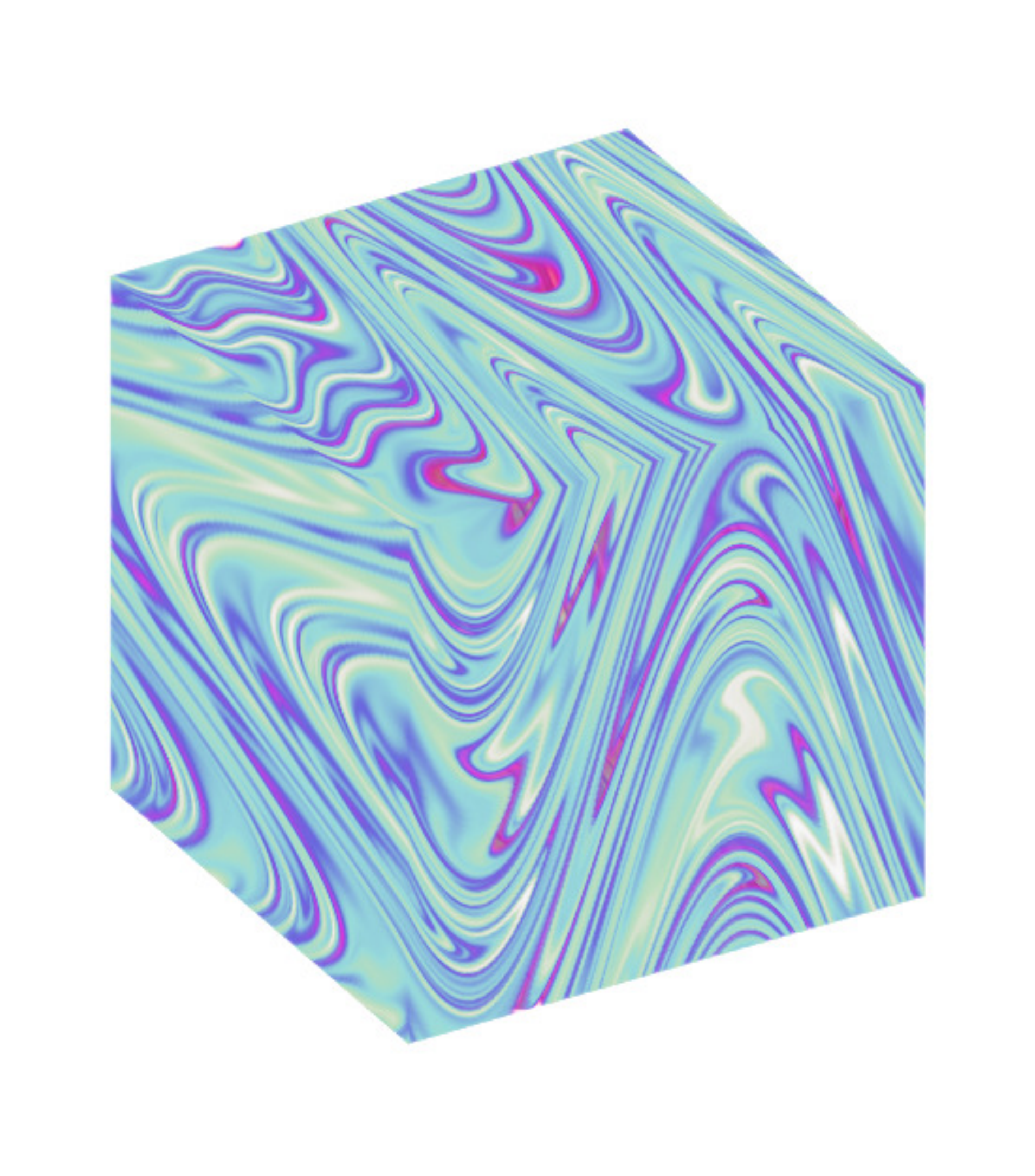}
 \end{minipage}
\begin{minipage}{5.175cm}
(b)
\includegraphics[scale=0.325,clip=true,trim=50 50 50 50]{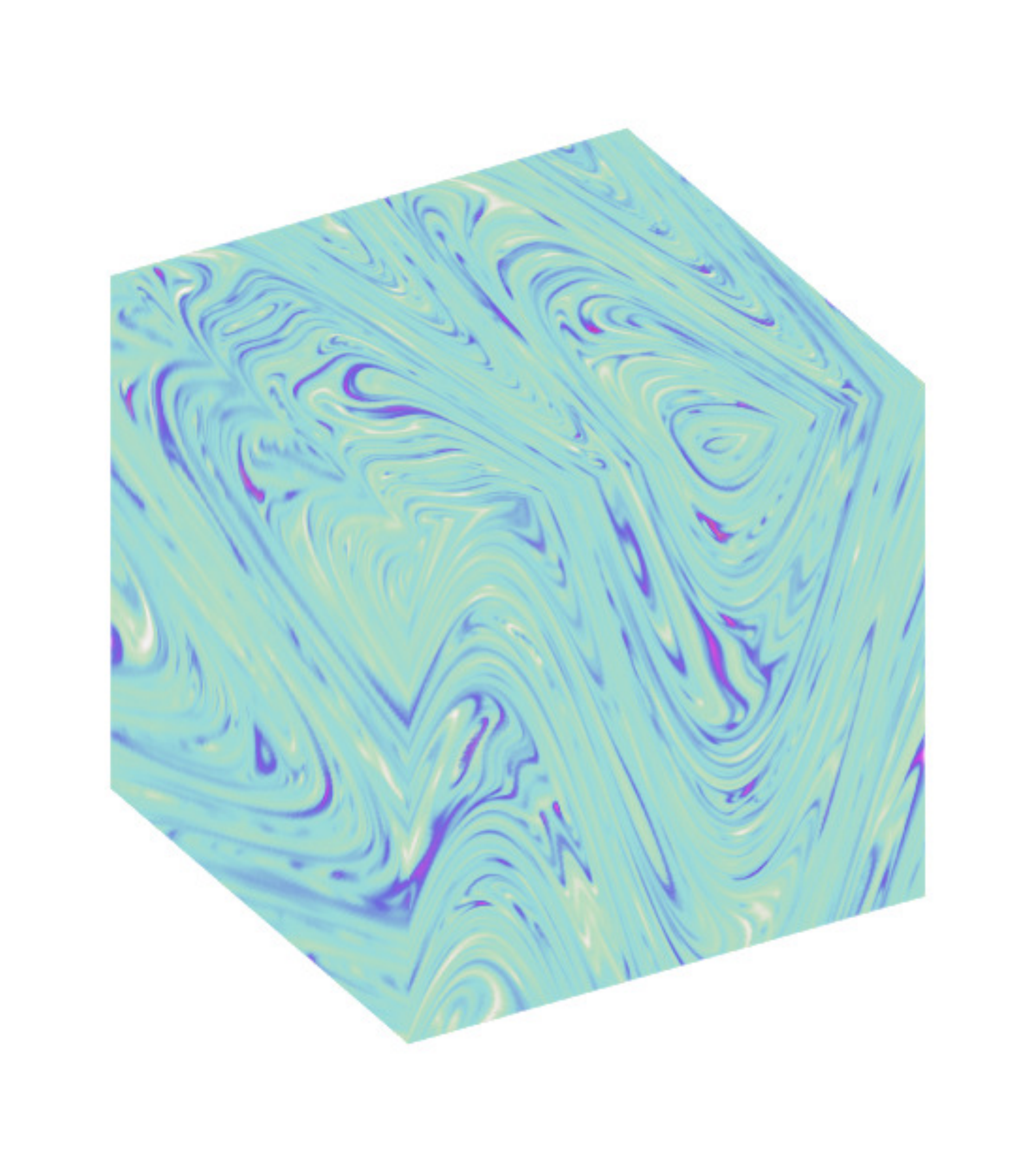}
 \end{minipage}
 \begin{minipage}{5.175cm}
(c)
\includegraphics[scale=0.325,clip=true,trim=50 50 50 50]{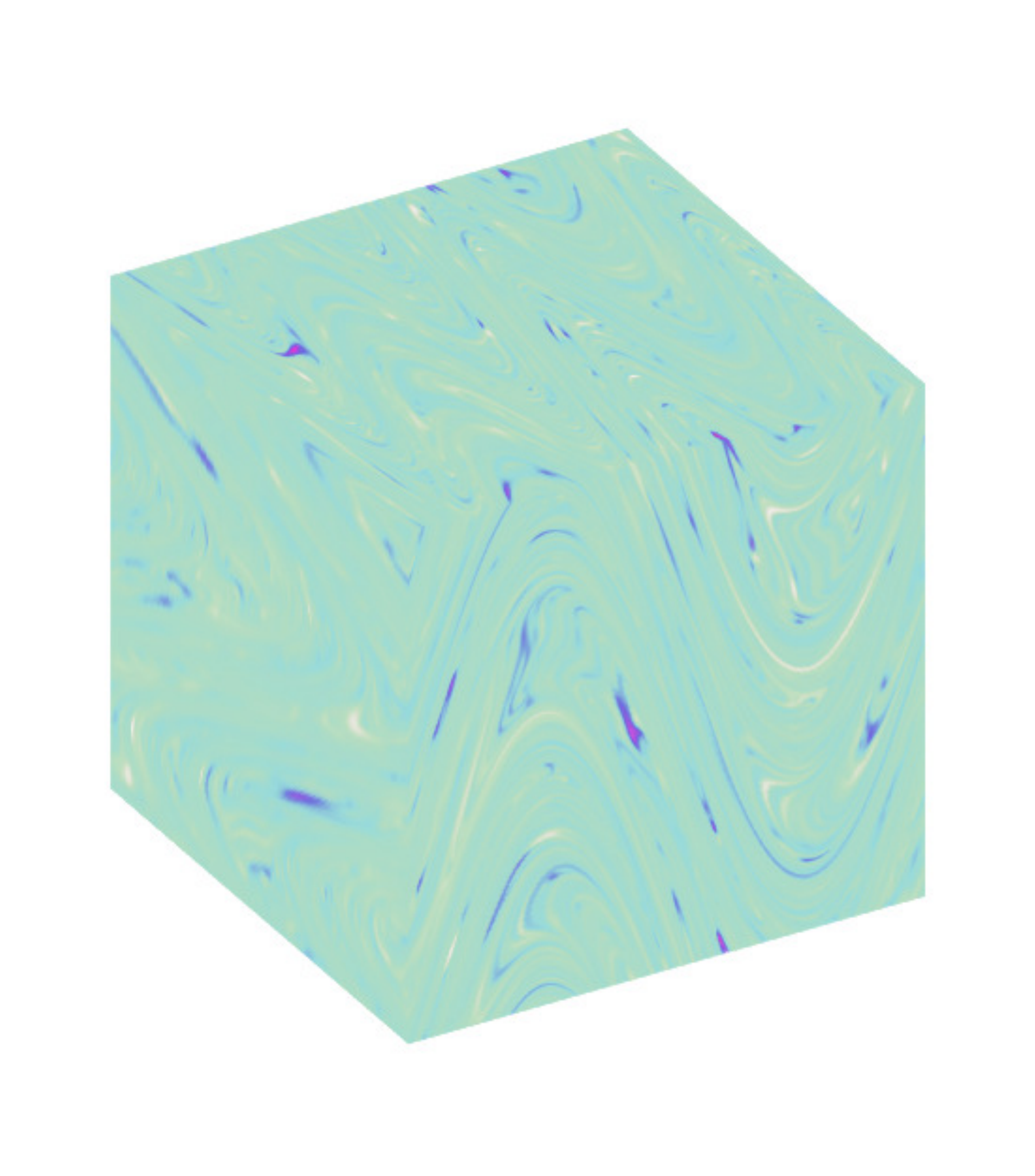}
 \end{minipage}

 \caption{(Colour online) Concentration field generated by the forward map \eqn{sine3d} with $\kappa=10^{-4}$: (a)
   $n=2$; (b) $n=4$; (c) $n=6$. Note the presence of filamentary sheets caused by expansion in two directions and contraction in one direction.}
\label{fig:tracer-P1-fwd}
\end{figure}

 \begin{figure}[t]
\begin{minipage}{5.175cm}
(a)
\includegraphics[scale=0.325,clip=true,trim=50 50 50 50]{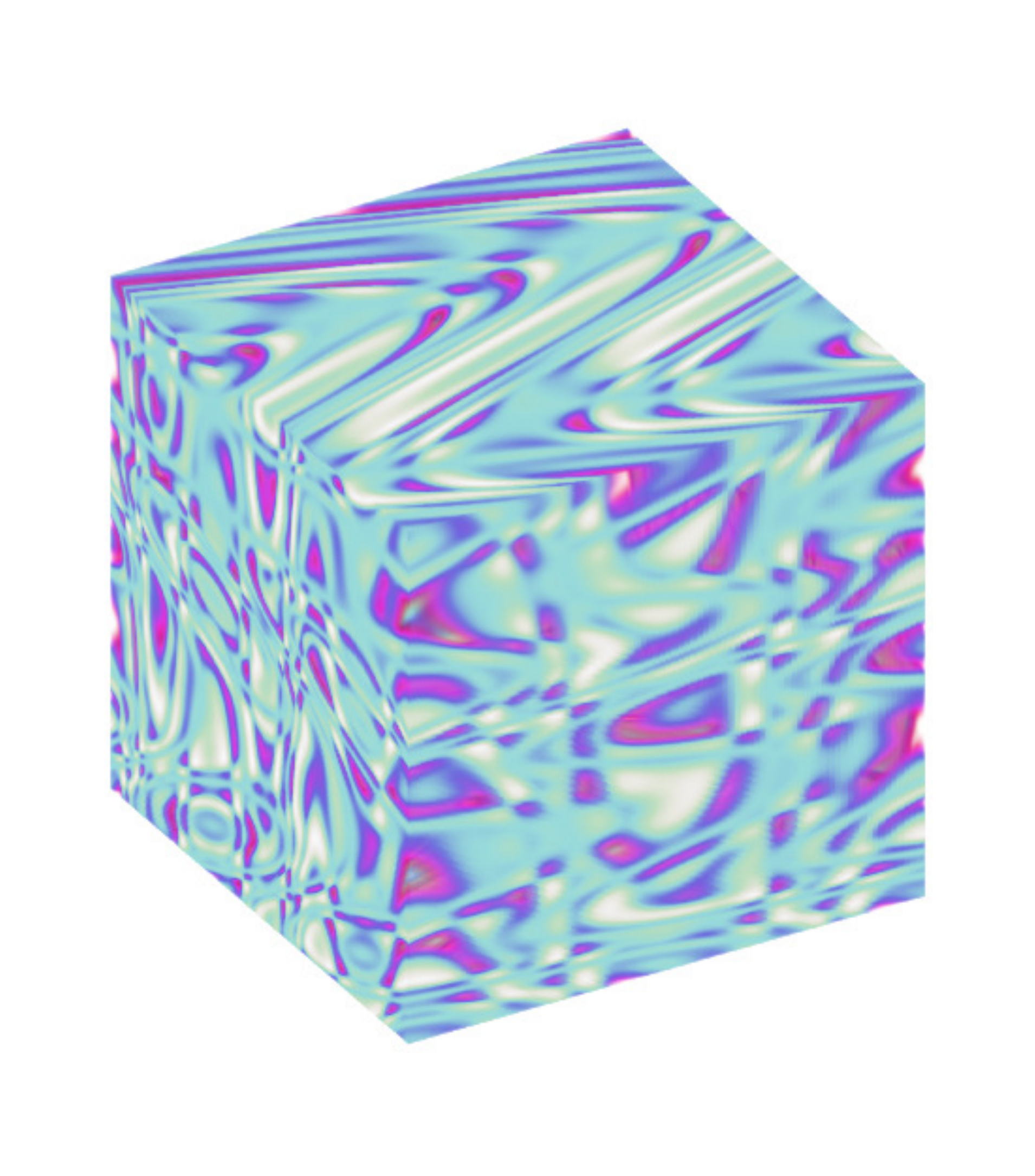}
 \end{minipage}
\begin{minipage}{5.175cm}
(b)
\includegraphics[scale=0.325,clip=true,trim=50 50 50 50]{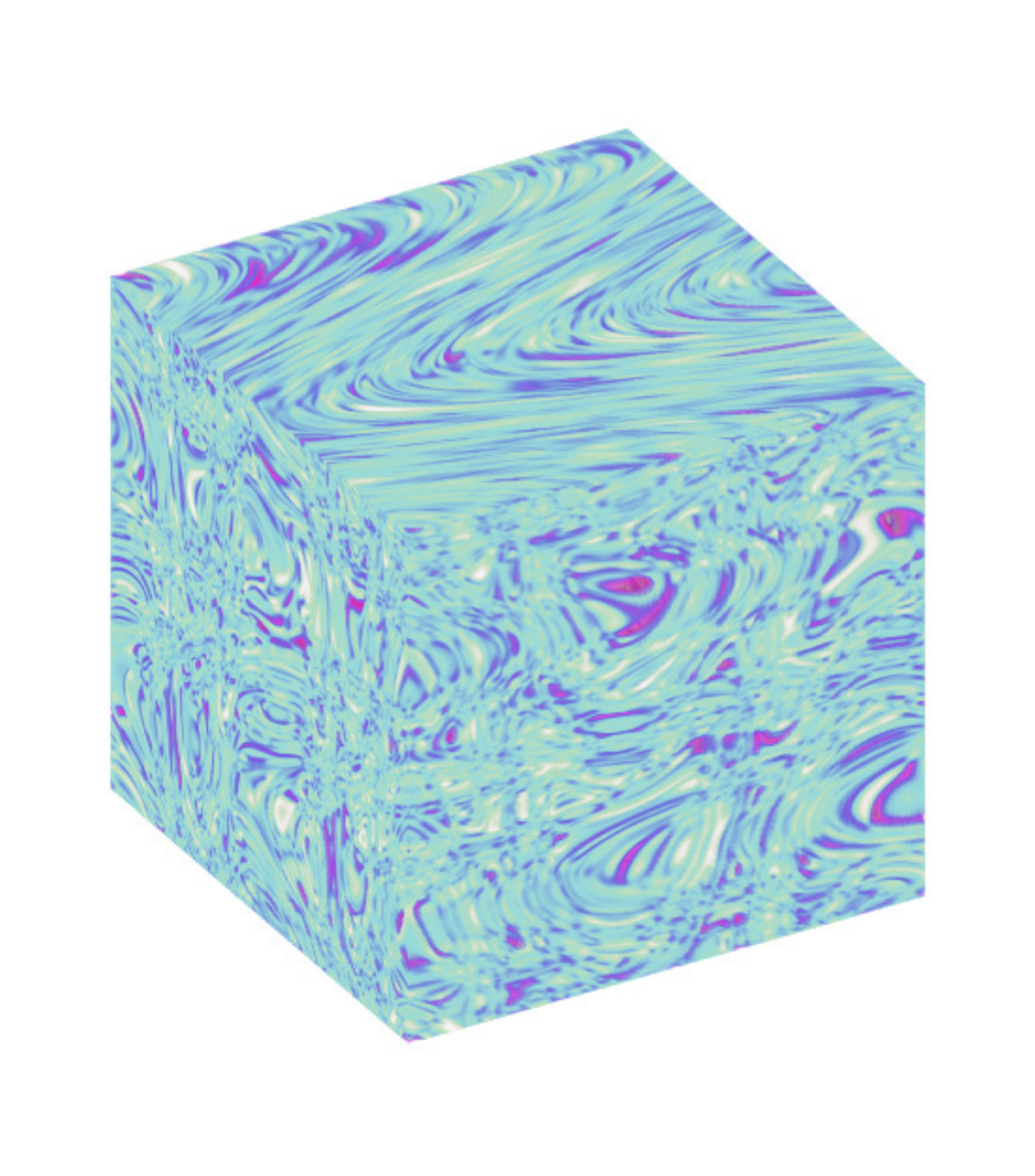}
 \end{minipage}
 \begin{minipage}{5.175cm}
(c)
 \includegraphics[scale=0.325,clip=true,trim=50 50 50 50]{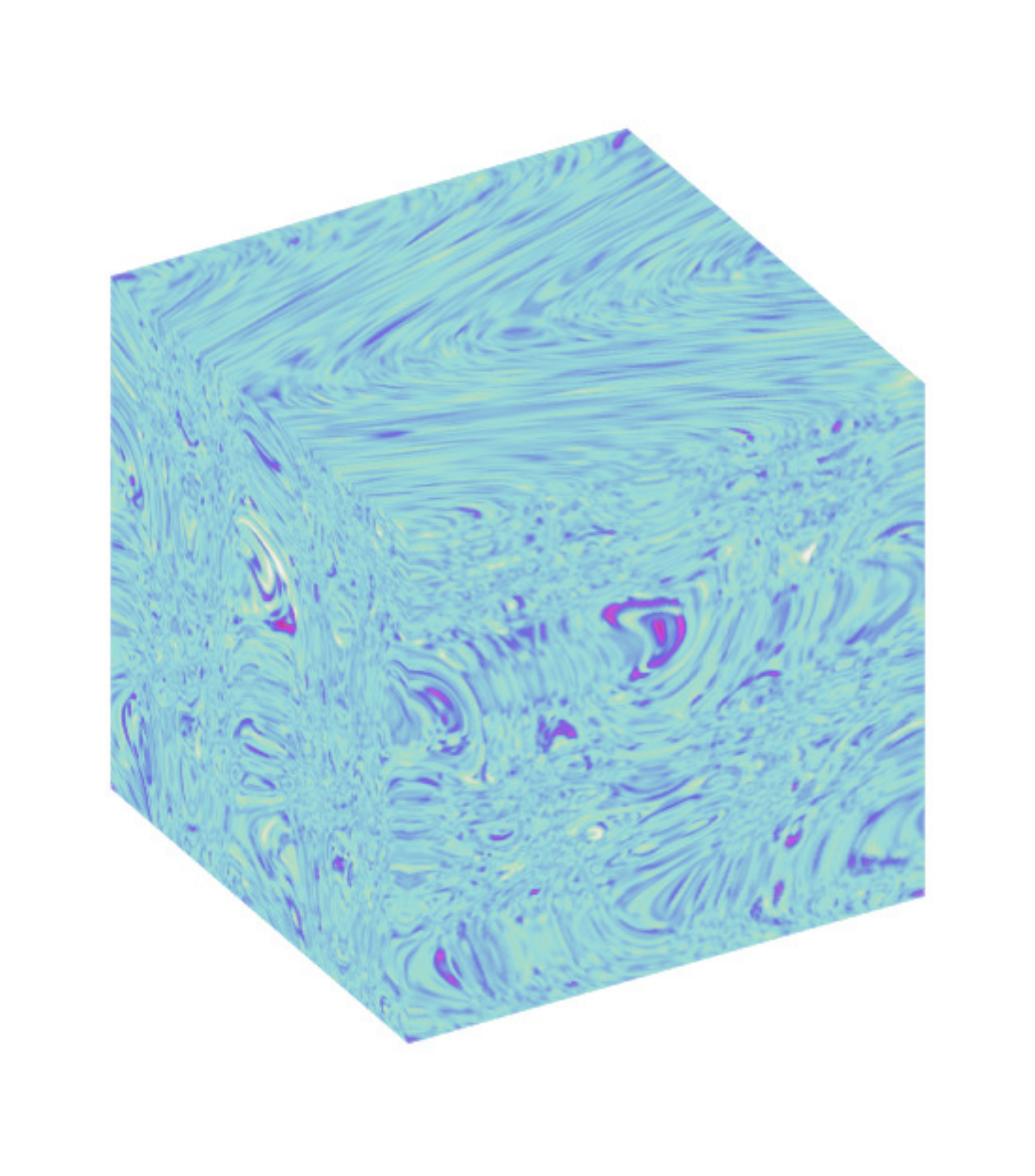}
 \end{minipage}
 \caption{(Colour online) Concentration field generated by the inverse map
   \eqn{sine3dinv} with $\kappa=10^{-4}$: (a) $n=2$; (b) $n=4$; (c) $n=6$. Note the presence of thin ``needles'' caused by expansion in one direction and contraction in two directions. }
\label{fig:tracer-P1-inv}
\end{figure}

The spatial structure of the decaying scalar is illustrated by Figs.\
\ref{fig:tracer-P1-fwd} and \ref{fig:tracer-P1-inv}, which show
volume-rendered concentration fields for the forward and inverse maps,
respectively. The plots use the normalised concentration variable
\begin{equation}
\theta = \frac{C}{\sigma},
\end{equation}
where $\sigma$ is the the standard deviation defined by \eqn{vardef}.
This provides a numerical approximation to the strange eigenmode
structure $D(\bx,t)$ defined in \eqn{adle}. To facilitate comparison,
the colour scale is renormalised in each panel, i.e., it spans the
respective minimum and maximum values.

For the forward map, there is a strong resemblance to the
two-dimensional case: large-scale filaments dominate, though in the
present case they are manifested as two-dimensional surfaces. This, of
course, results from the presence of two expanding and one contracting
directions. 

For the inverse map, the picture is quite different. In place of the
large-scale filamentary sheets, there are now thin ``needles'', which
arise from the presence of two contracting directions. This is most
obvious for $n=4$.  Note that these needles are approximately parallel
to the $(x,y)$-plane: the final advection step of each iteration
consists of a purely horizontal shear (see \eqn{sine3dinv}) which
creates elongated structures in the $(x,y)$-plane. The contrast
between Figs.\ \ref{fig:tracer-P1-fwd} and \ref{fig:tracer-P1-inv}
underscores the profound effect that differences in the stretching
properties of the forward and inverse maps --- in particular, the
number of positive Lyapunov exponents --- have on the structure of the
concentration field.  It is therefore remarkable that the variance
decay rate is insensitive to this. The physical explanation for this
result is that the variance for large $n$ is controlled by isolated
pockets of extreme concentration values; these pockets, which can be
readily identified in Figs.\ \ref{fig:tracer-P1-fwd}c and
\ref{fig:tracer-P1-inv}c and correspond to rare events of low
stretching \citep{hayn-v05}, have the same statistics in the forward
and inverse maps (cf.~Sec.~\ref{sec:decay}).

\section{Transition to global control: $P > 1$}
\label{sec:trans-glob-contr}

In this section we examine the transition between the locally
controlled and globally controlled regimes by analysing the scalar
decay in domains of size $2\pi P$, $P>1$. We focus on the forward map \eqn{sine3d}
and fix $\kappa=10^{-4}$ and $N=1024$. Simulations have been carried
out with the two initial conditions \eqn{ics_local} and
\eqn{ics_global}.

\subsection{Variance decay and concentration fields}

 \begin{figure}[t]
\begin{center}
\includegraphics[scale=0.42]{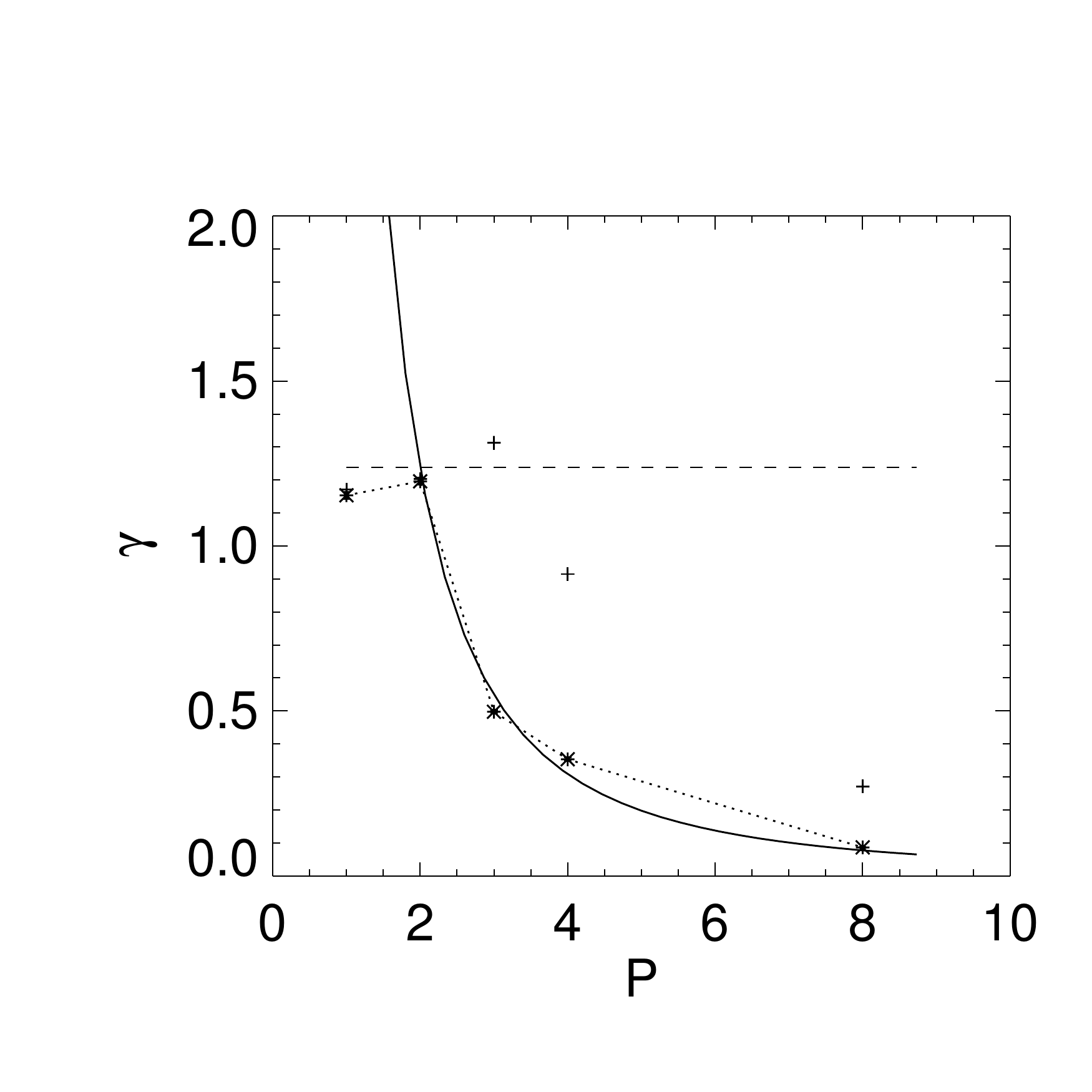}
\end{center}
\caption{Decay rate $\gamma$ versus $P$ for $\kappa=10^{-4}$. The
  numerical results obtained with the initial conditions
  \eqn{ics_local} [crosses, $+$] and \eqn{ics_global} [asterisks, $*$] are compared with
  the prediction from homogenisation theory $\gbar_\mathrm{global}$
  (long dashes) and with the prediction 
  $\gbar_\mathrm{local}$ for the locally controlled regime (solid line).}
\label{fig:vardecay-P}
\end{figure}

\begin{figure}[t]
 \begin{center}
 \begin{minipage}{5.175cm}
(a)
\includegraphics[scale=0.325,clip=true,trim=50 50 50 50]{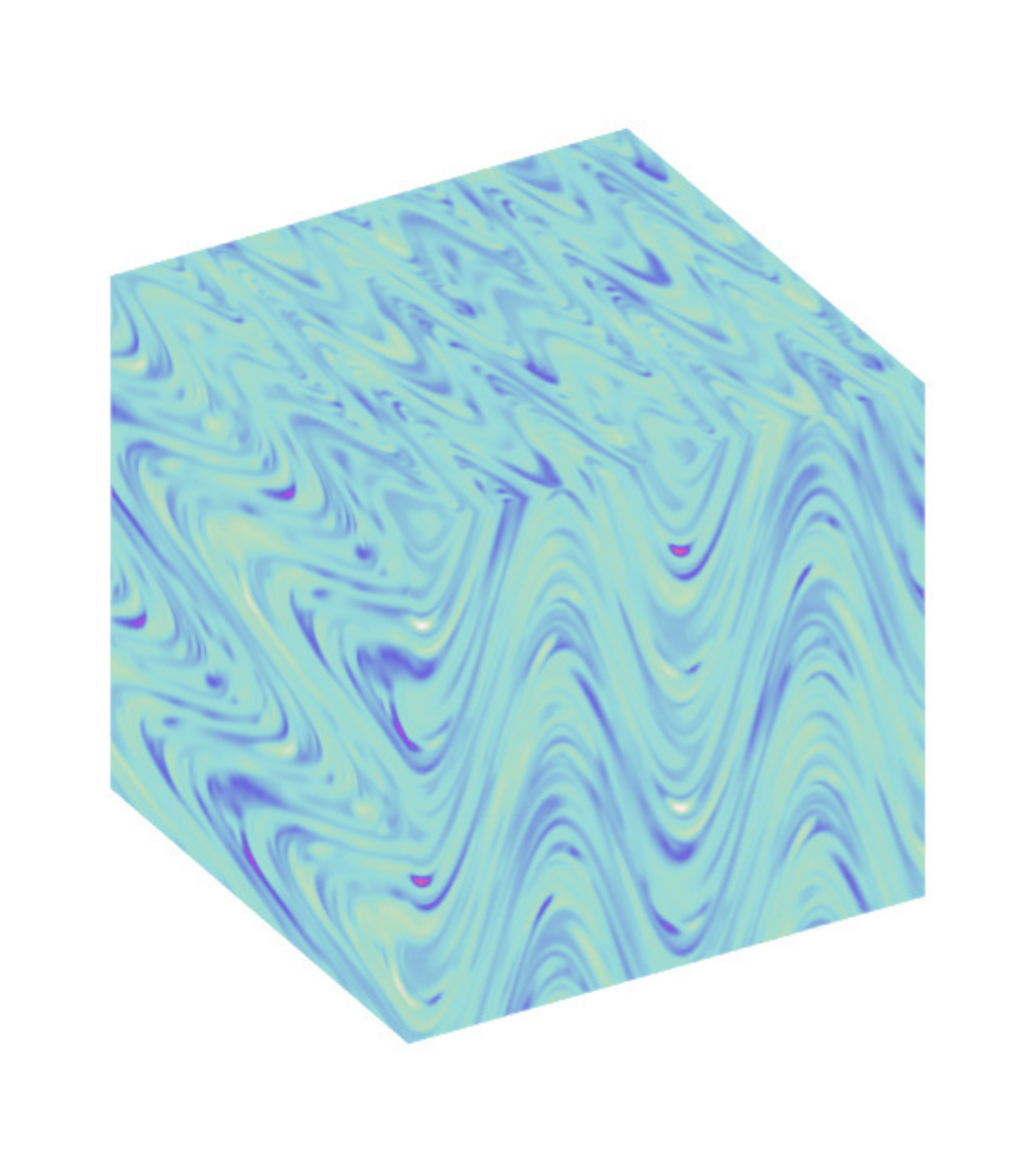}
\end{minipage} \hspace{3cm}
 \begin{minipage}{5.175cm}
(b) \includegraphics[scale=0.325,clip=true,trim=50 50 50 50]{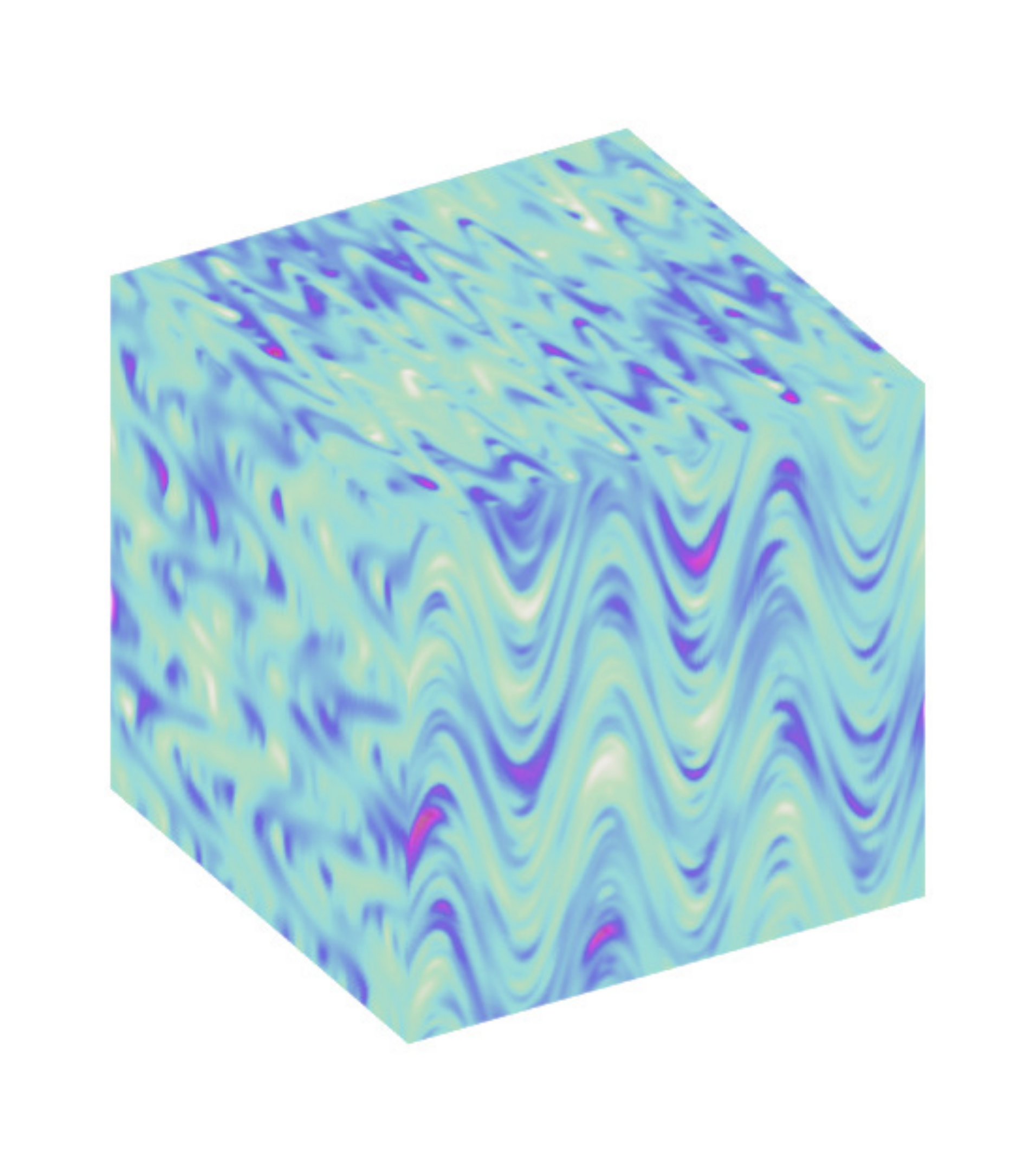} 
 \end{minipage}

 \begin{minipage}{5.175cm}
 (c)
 \includegraphics[scale=0.325,clip=true,trim=50 50 50 50]{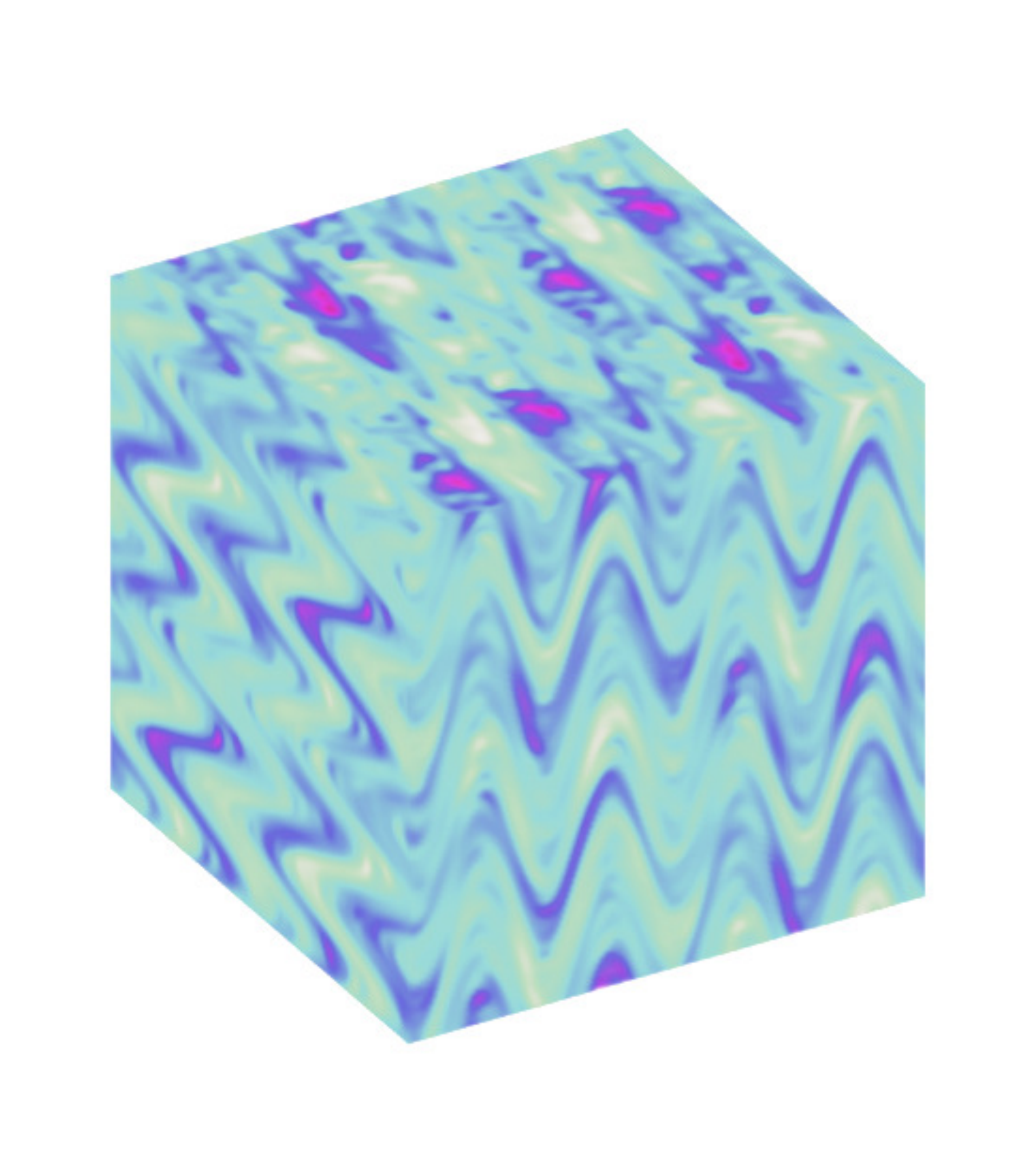}
\end{minipage} \hspace{3cm}\begin{minipage}{5.175cm}
(d)
\includegraphics[scale=0.325,clip=true,trim=50 50 50 50]{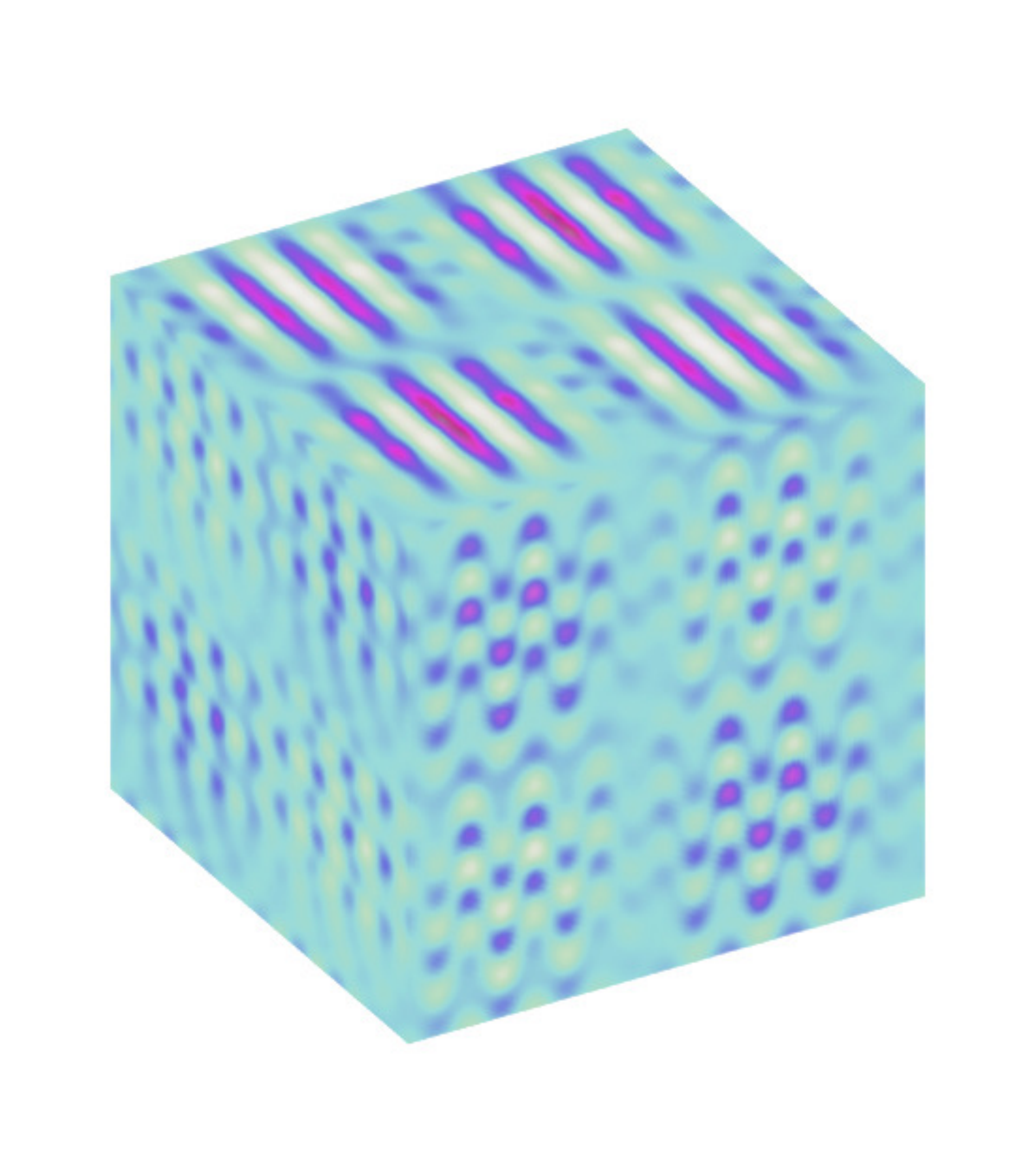}
 \end{minipage}


\caption{(Colour online) Concentration fields for $P\ge2$, $n=6$ and $\kappa=\ttexp{-4}$ with initial conditions \eqn{ics_local}: (a) $P=2$;
  (b) $P=3$; (c) $P=4$;  (d) $P=8$. Note the transition from a filamentary to a
periodic appearance as $P$ is increased.}

\label{fig:tracer-global}
\end{center}
\end{figure}

Fig.~\ref{fig:vardecay-P} summarises the variance decay results by
showing the decay rate $\gamma$ as a function of $P$. The numerical
results are compared with the asymptotic predictions \eqn{localdecay}
and \eqn{globaldecay} for the locally and globally controlled regimes.
The existence of the two distinct regimes is clear from the figure,
with the transition taking place around $P=3$. The structure of the
scalar field for $P=2$ (see Fig.\ \ref{fig:tracer-global}(a)), which
is very similar to the structure for $P=1$, suggests that $P=2$ is in
the locally controlled regime. This is consistent with the argument
outlined in Sec.~\ref{sec:theor-pred} which predicts that $P=3$
corresponds to the smallest domain for global control. 


The decay rates found for $P \ge 3$ with the initial conditions
\eqn{ics_local} are much larger than the approximation to
$\gbar_\mathrm{global}$ provided by homogenisation theory in
\eqn{globaldecay}. We attribute this discrepancy to finite-time or symmetry
effects: in the homogenisation scenario, the decaying concentration
field has a structure close to that of one of the gravest modes of the
diffusion operator, that is $\sin(x/P)$, $\sin(y/P)$ or $\sin(z/P)$, depending
on details of the initial conditions. However, in our simulations with
the initial conditions \eqn{ics_local}, the concentration fields are
very different (see Figs.\ \ref{fig:tracer-global}(b)--(d)). This can
be understood as a finite-time effect. As the diffusive approximation
of homogenisation theory indicates, the relative amplitudes of the
Fourier modes depend on the initial conditions; therefore the high
modes can dominate for some finite time if their initial amplitudes
are substantially larger than the amplitudes of the gravest modes.
This appears to be the case with the initial conditions
\eqn{ics_local}, which would explain why the prediction
\eqn{globaldecay}, although valid for $n \gg 1$, is not realised in
our simulations (which are limited to $n \le 30$). Furthermore, for $P \ge 2$ even, the dynamics preserves the symmetries $C(x+P\pi,y,z)=-C(x,y,z)$, $C(x,y+P\pi,z)=-C(x,y,z)$ and $C(x,y,z+P\pi)=-C(x,y,z)$ of the initial conditions \eqn{ics_local} (see Figs.\ \ref{fig:tracer-global}(a),(c),(d)); however, these symmetries are not compatible with the gravest-mode structure.

 \begin{figure}[t]
 \begin{minipage}{5.175cm}
(a)
 \includegraphics[scale=0.32,clip=true,trim=50 50 50 50]{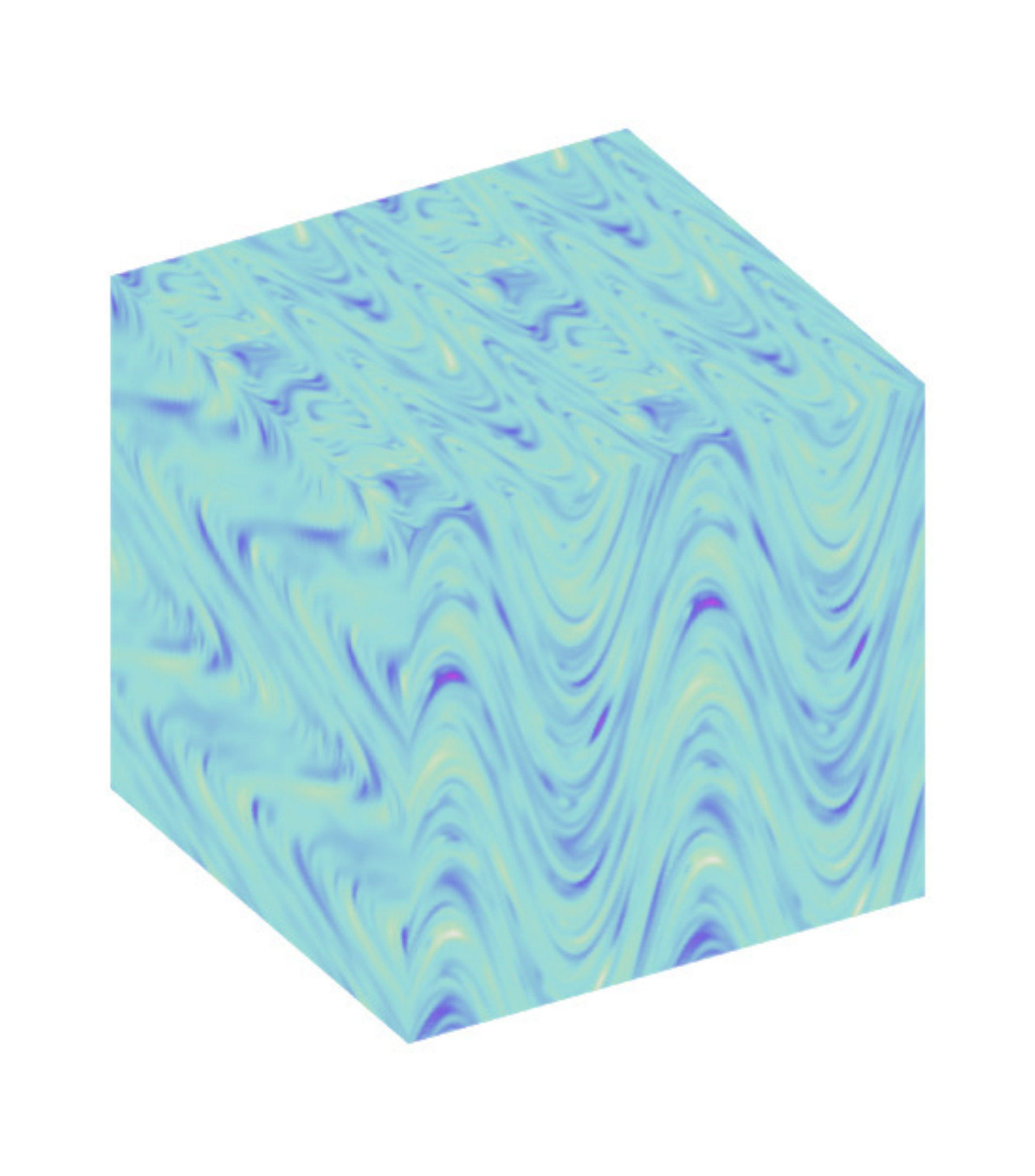}
 \end{minipage}
 \begin{minipage}{5.175cm}
(b)
 \includegraphics[scale=0.32,clip=true,trim=50 50 50 50]{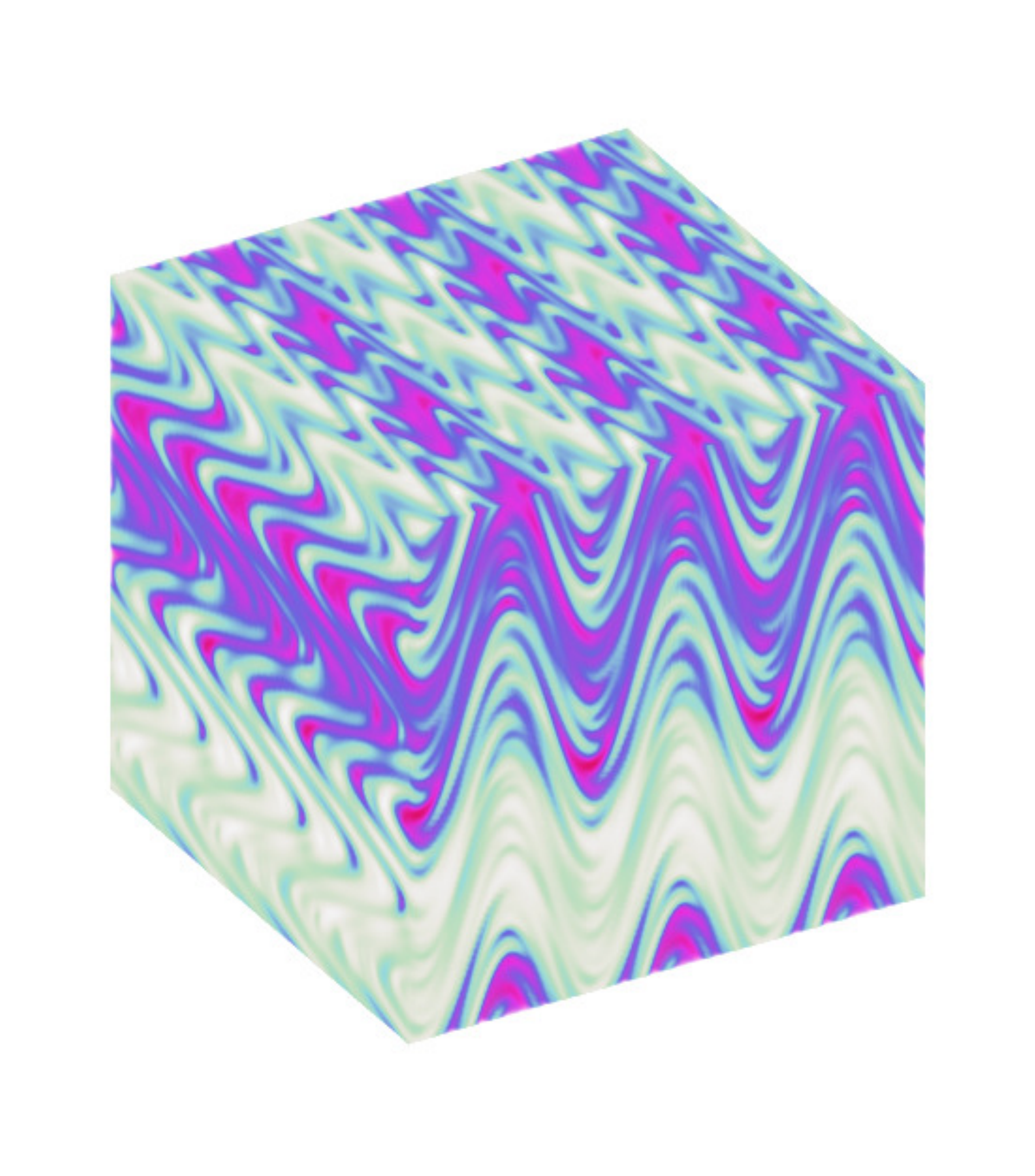}
 \end{minipage}
 \begin{minipage}{5.175cm}
(c)
 \includegraphics[scale=0.32,clip=true,trim=50 50 50 50]{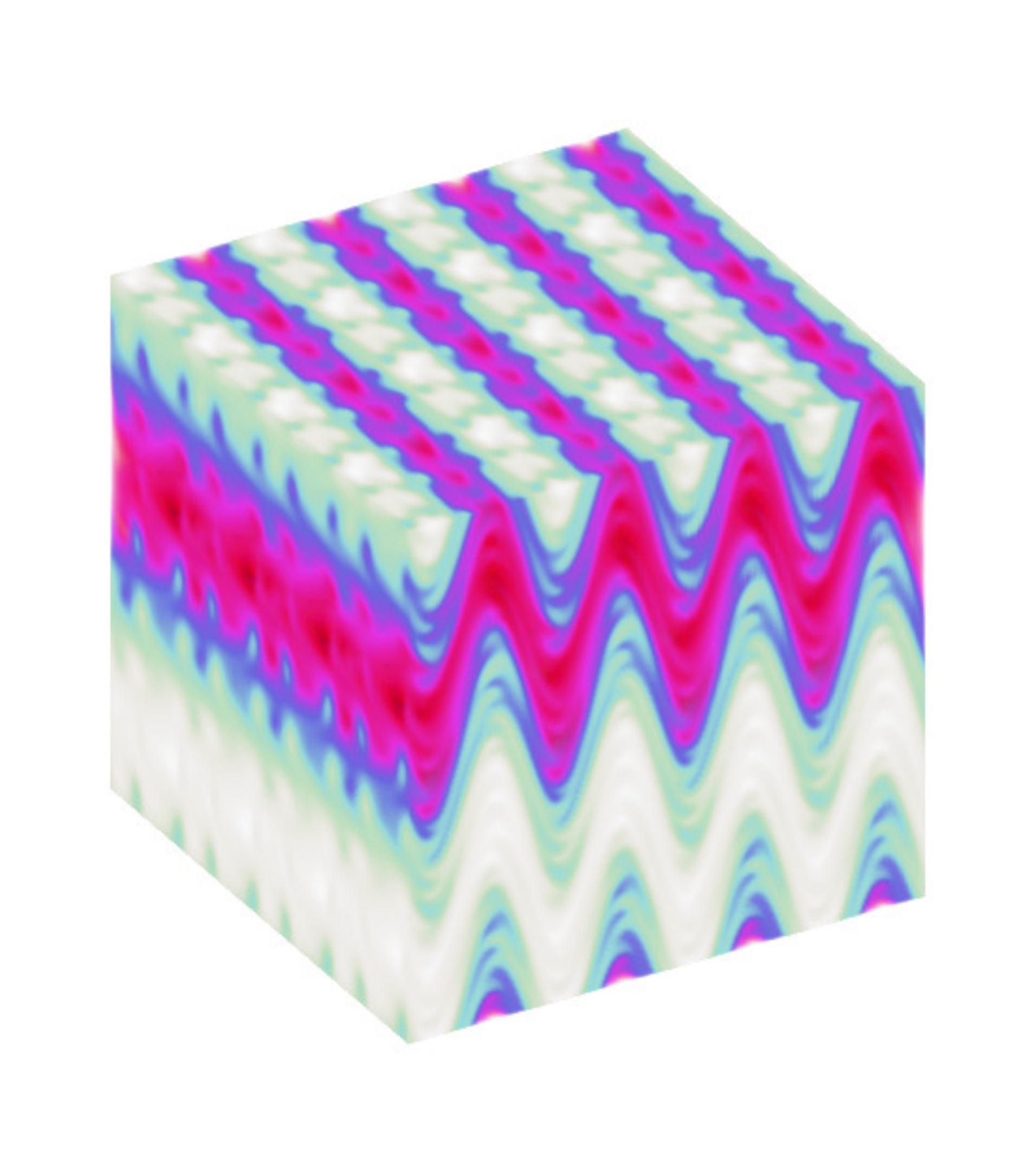}
 \end{minipage} 
 \caption{(Colour online) Concentration fields for $\kappa=\ttexp{-4}$ with initial conditions \eqn{ics_global}: (a) $P=2$; (b) $P=3$; (b) $P=4$. As $P$ increases, the structure of the concentration approaches the structure of one of the gravest modes of the diffusion operator, here $\sin z$.}

\label{fig:tracer-sinz}
\end{figure}

To confirm that the discrepancy between simulated and predicted decay rates results from finite-time effects and the special initial conditions, we have repeated the simulations using the gravest
mode \eqn{ics_global} as initial condition. In this case, the observed
decay rates closely match those predicted by homogenisation theory.
The scalar fields, displayed in Fig.\ \ref{fig:tracer-sinz}, have the
expected structure of a distorted gravest mode for $P \ge 3$, with the
distortions reducing as $P$ increases. Comparison between the scalar
fields for $P=2$ and $P=3$ further supports our claim of local control
for $P=2$.

\subsection{Statistical moments and pdfs}
\label{sec:stat-moments-pdfs}

 \begin{figure}[t]
 \begin{center}
 \begin{tabular}{cc}
 (a) & (b) \\
 \includegraphics[scale=0.4]{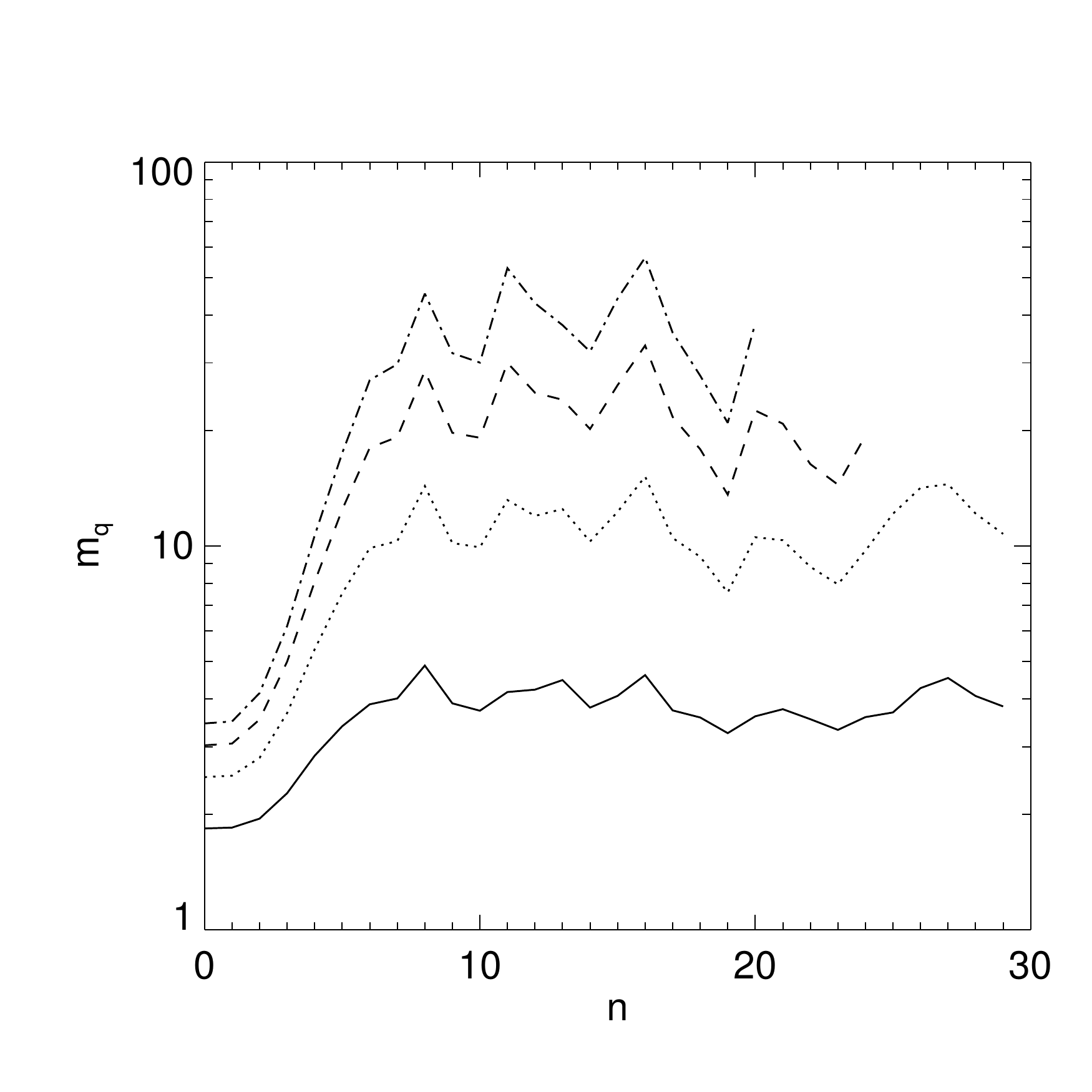} &
 \includegraphics[scale=0.4]{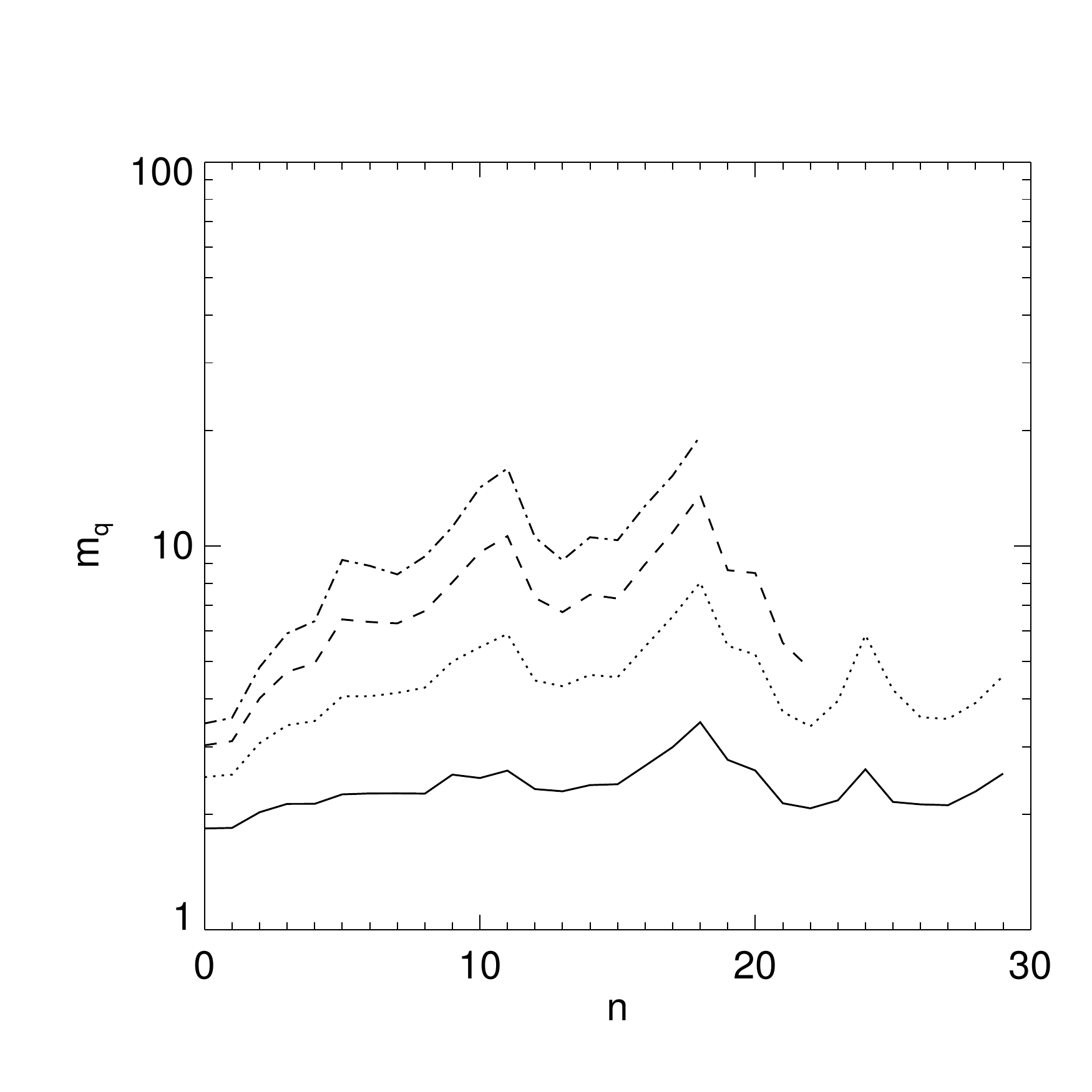} \\
 (c) & (d) \\
 \includegraphics[scale=0.4]{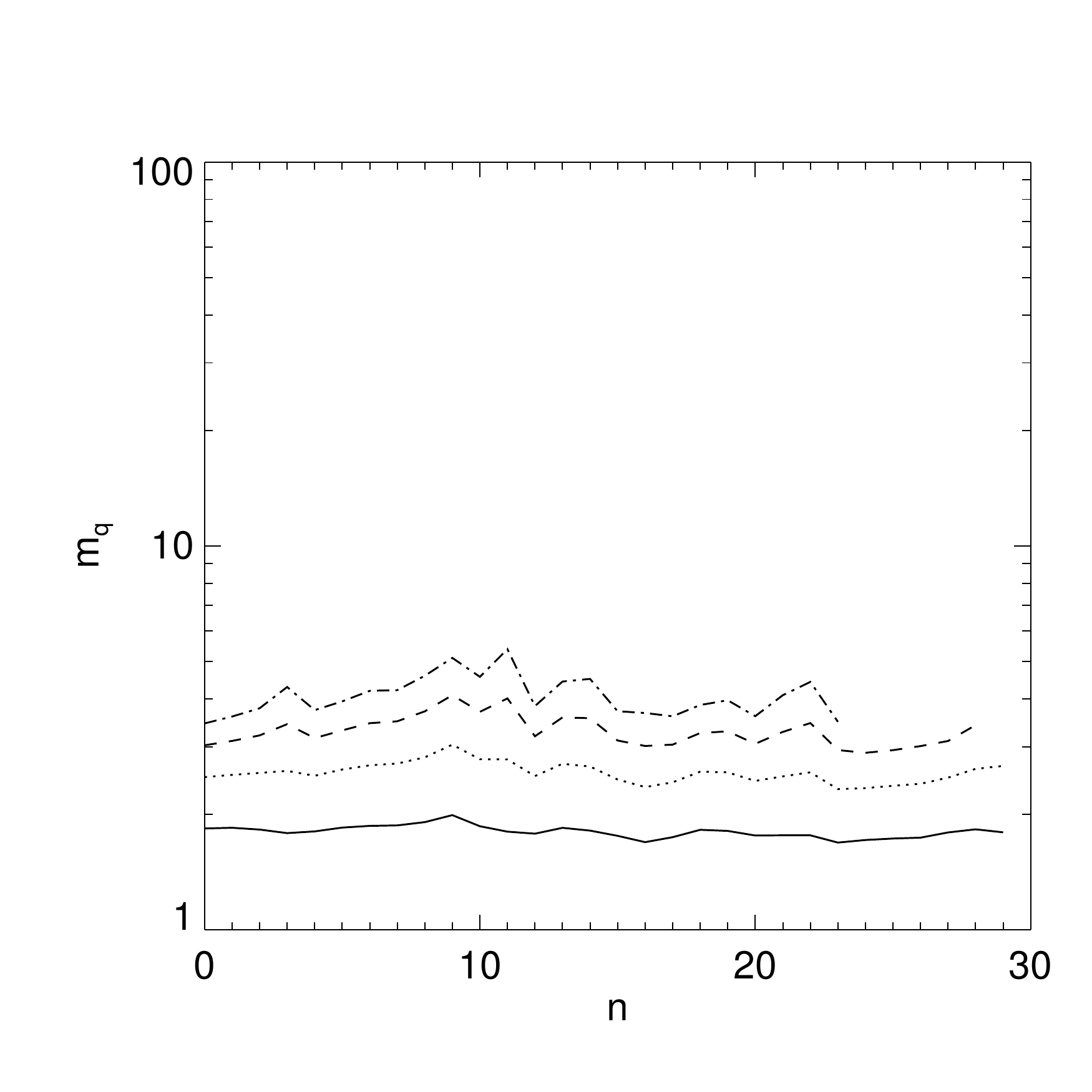} &
\includegraphics[scale=0.4]{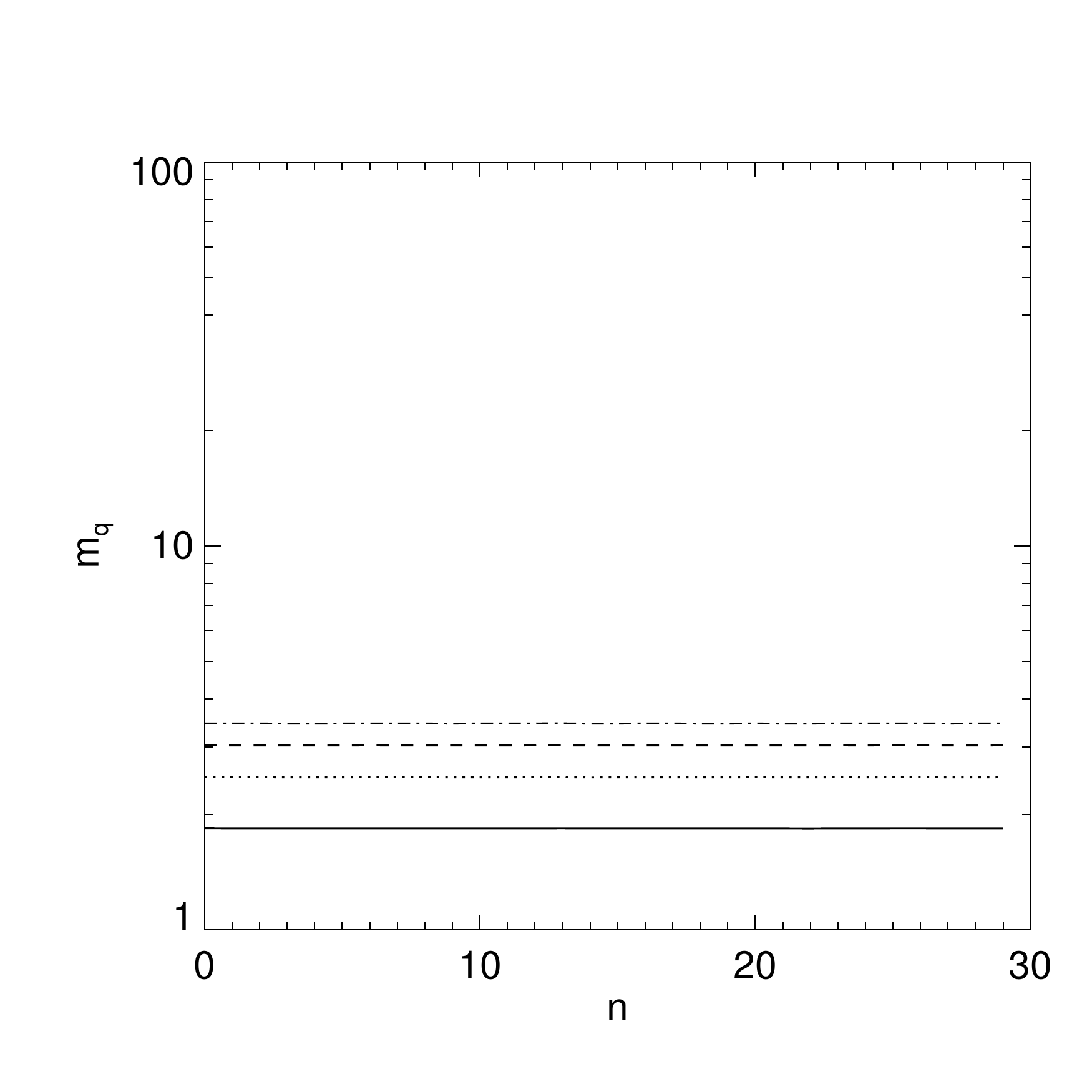}
 \end{tabular} 
  \caption{Time evolution of the moments, $m_q$ vs $n$, for the forward map
   with $\kappa=10^{-4}$: (a) $P=1$; (b) $P=2$; (c) $P=4$; (d) $P=8$. }

\label{fig:moments}
\end{center}
\end{figure}

 \begin{figure}[t]
 \begin{center}
 \begin{tabular}{cc}
 (a) & (b) \\
 \includegraphics[scale=0.4]{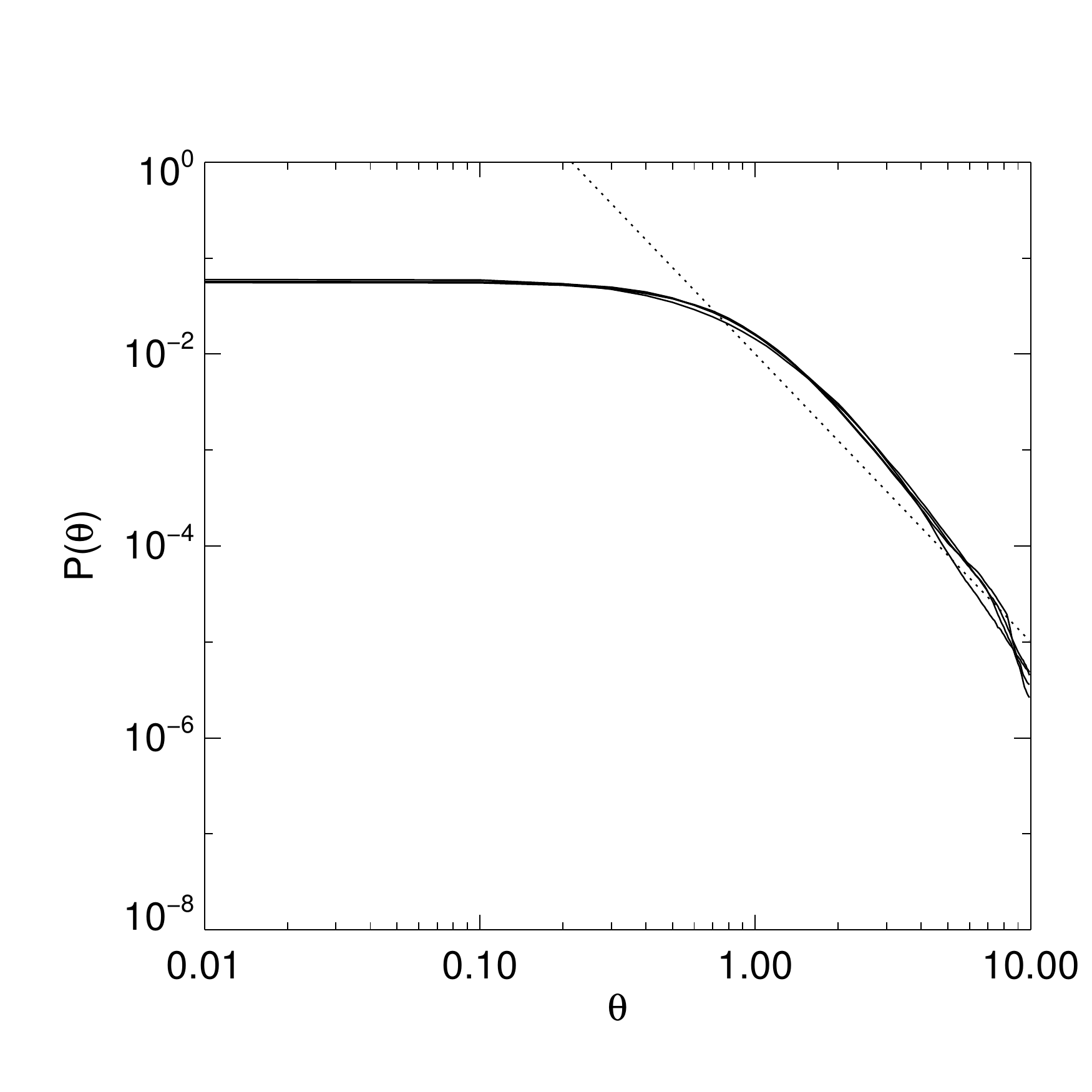} &
 \includegraphics[scale=0.4]{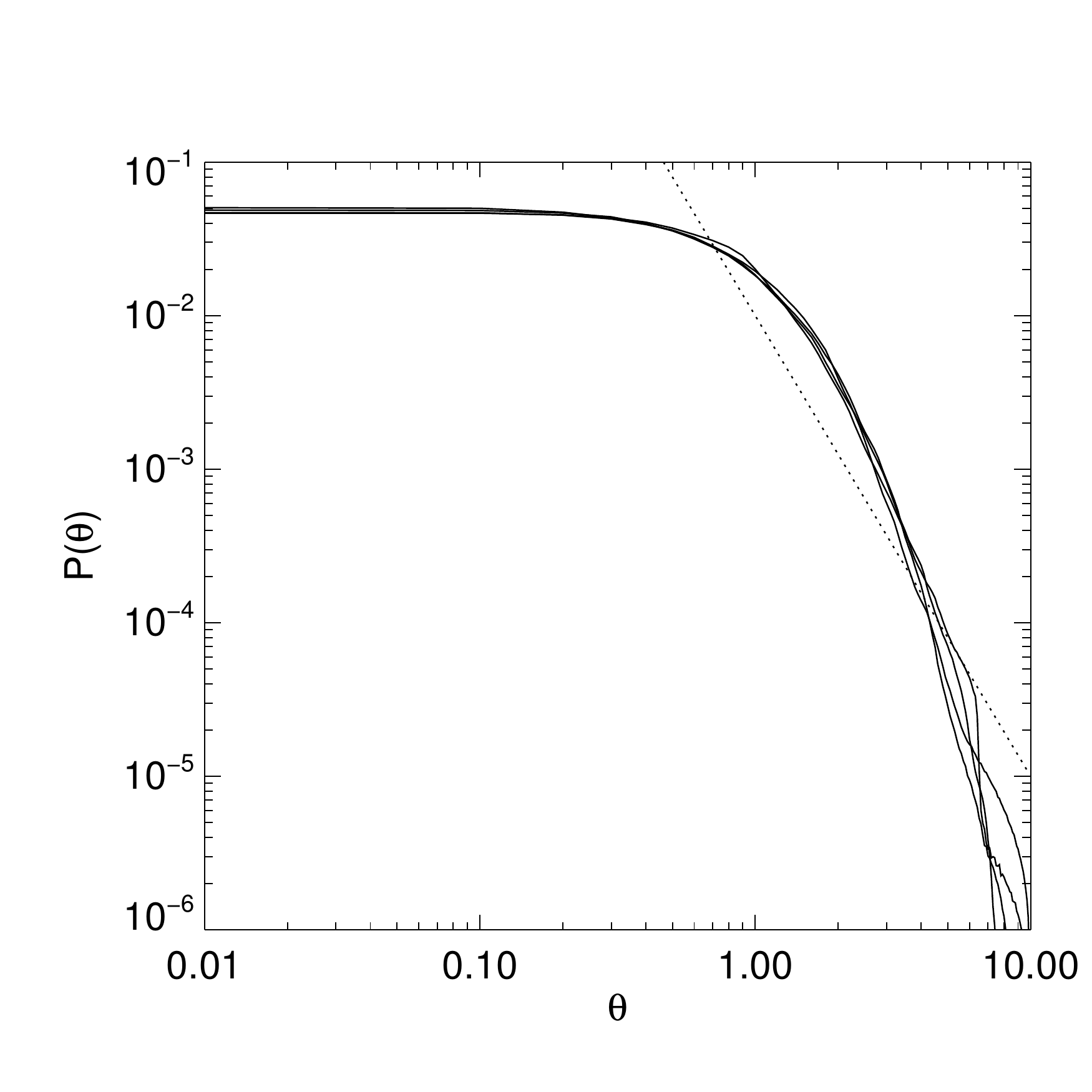} \\
 (c) & (d) \\
 \includegraphics[scale=0.4]{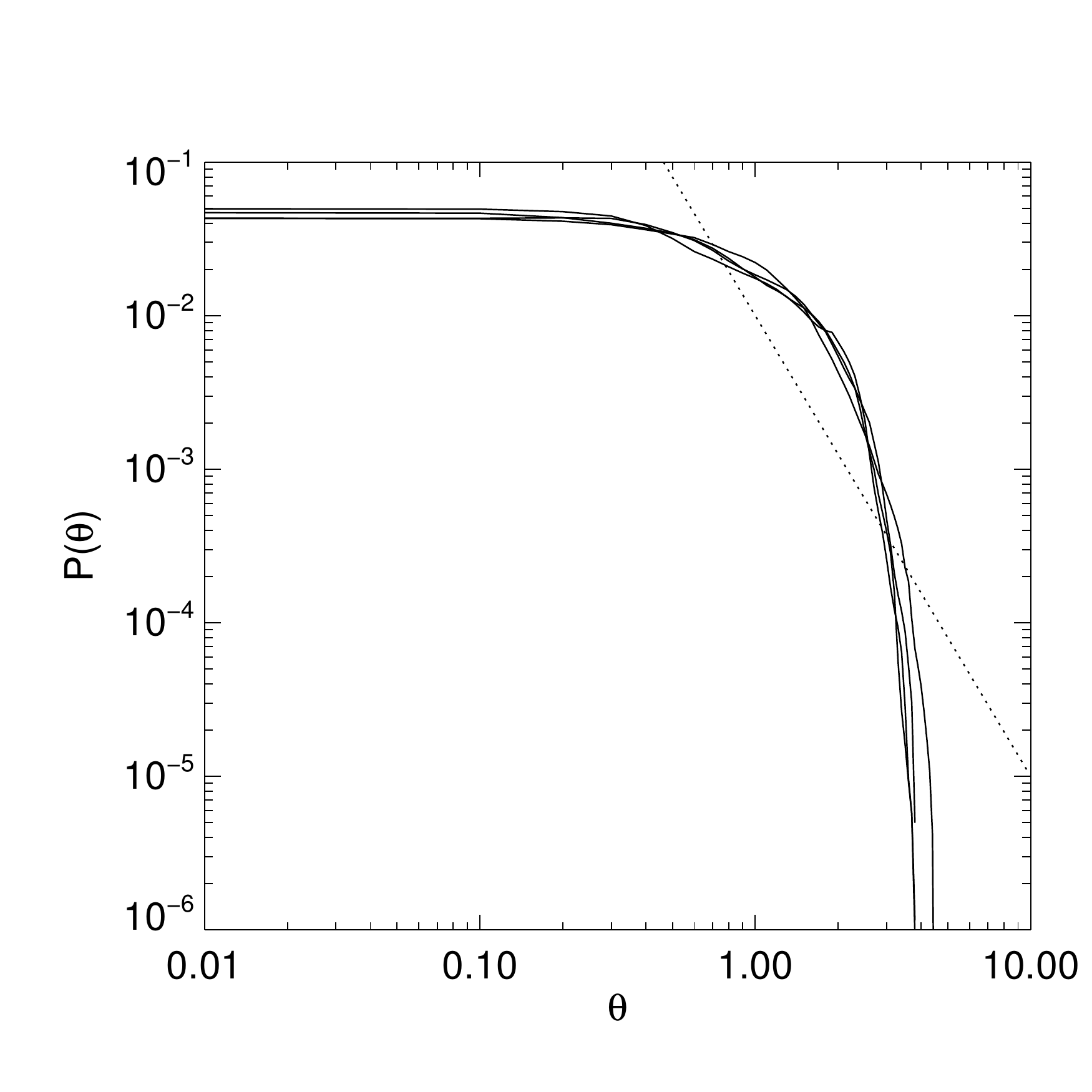} &
\includegraphics[scale=0.4]{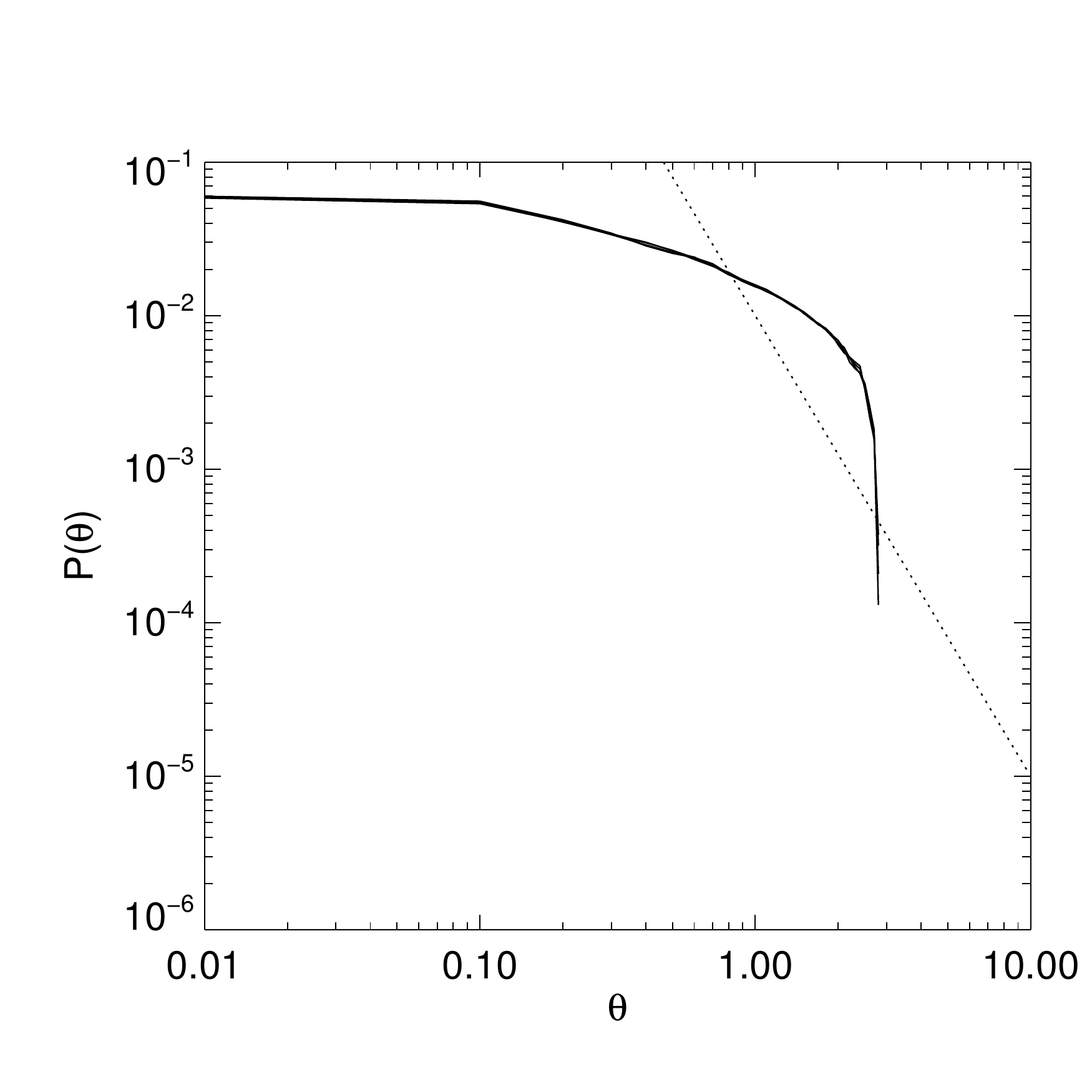}
 \end{tabular}


 \caption{$p(\theta)$ vs $\theta$ for the runs analysed in
   Fig.~\ref{fig:moments}: (a) $P=1$; (b) $P=2$; (c) $P=4$; (d) $P=8$.
   The solid lines correspond to iterations $n=5,10,20,30$; the
   reference slope $|\theta^{-3}|$ is plotted with a dotted line. In
   the global regime (panels c and d), the tails are much shorter. The
   pdfs are obtained by binning $\theta\in[-10,10]$ with 200 bins of
   uniform width.}

\label{fig:pdfs}
\end{center}
\end{figure}

An alternative view of the transition from local to global control is
provided by statistical moments and pdfs of the concentration field. 
Normalised moments are defined by 
\begin{equation}
m_q = \frac{1}{\sigma}\left(\int C ^q dx \right)^{2/q}.
\end{equation}
According to the strange-eigenmode expression \eqn{adle}, they should be stationary random functions. This was confirmed in numerical simulations of the 2-D sine map, which indicate that the $m_q$ are
approximately constant \citep{fere-hayn}. This is at odds with the predictions of \citep{balk-foux} that the $m_q$ depend on $q$ and
increase exponentially with time.

Fig.~\ref{fig:moments} plots $m_q$ against $n$ for various values of
$P$ and $q=4,6,8,10$. Focusing on $P=1$, the $m_q$ increase rapidly
(approximately exponentially) on short timescales before levelling
off, in close analogy to the two-dimensional results (see Fig.~3 of
\citep{fere-hayn}).  After the initial transient period, the
$m_q$ appear roughly independent of $n$, consistent with \eqn{adle}
and the existence of a statistical equilibrium established by
stretching and diffusion.  For $P=2$ the picture is similar, though
the initial growth of the $m_q$ is smaller. For $P=4$ and $P=8$, by
contrast, there is hardly any growth at all. This is consistent with
the tracer behaviour being controlled by an effective diffusion for
large $P$.

With regards to the concentration pdfs, Fereday and Haynes \citep{fere-hayn} argued that the tails should scale like $|\theta|^{-\beta}$ in the limit $\kappa \to 0$, where 
\begin{equation}
\beta \sim 1 + 2g(0)/\gamma.
\end{equation}
This implies that $\beta \sim 3$ in the locally controlled regime,
since $\gamma \sim \bar \gamma_\mathrm{local} \sim - \ell(q_*) =
g(0)$; on the other hand, $\beta > 3$ in the globally controlled
regime, since $\bar \gamma_\mathrm{global} < \bar
\gamma_\mathrm{local}$. This prediction was verified in the 2D
case \citep{fere-hayn,hayn-v05}. Here we examine this prediction for
the forward 3D map \eqn{sine3d}.

Fig.~\ref{fig:pdfs} shows log-log plots of the pdf $p(\theta)$ for the
simulations analysed in Fig.~\ref{fig:moments}. The solid lines
correspond to pdfs evaluated at $n=5,10,20,30$, the dotted lines to
$|\theta|^{-3}$. We expect the spectral slope of the tails to change
as $P$ increases and we move from a local to global regime. For $P=1$
the slope is slightly steeper than the theoretical prediction, but the
algebraic decay is clear enough. The discrepancy may be related to the
finiteness of $\kappa$.  There is similar behaviour for $P=2$. For
$P=4$ and $P=8$, however, the prediction of algebraic tails breaks
down. Here the pdfs decays rapidly for $|\theta| > 1$. This is
consistent with the concentration field being controlled by an
effective diffusion in the homogenisation regime $P \gg 1$.

\section{Discussion}
\label{sec:discussion}

 \begin{figure}[t]
 \begin{center}
 \includegraphics[scale=0.32]{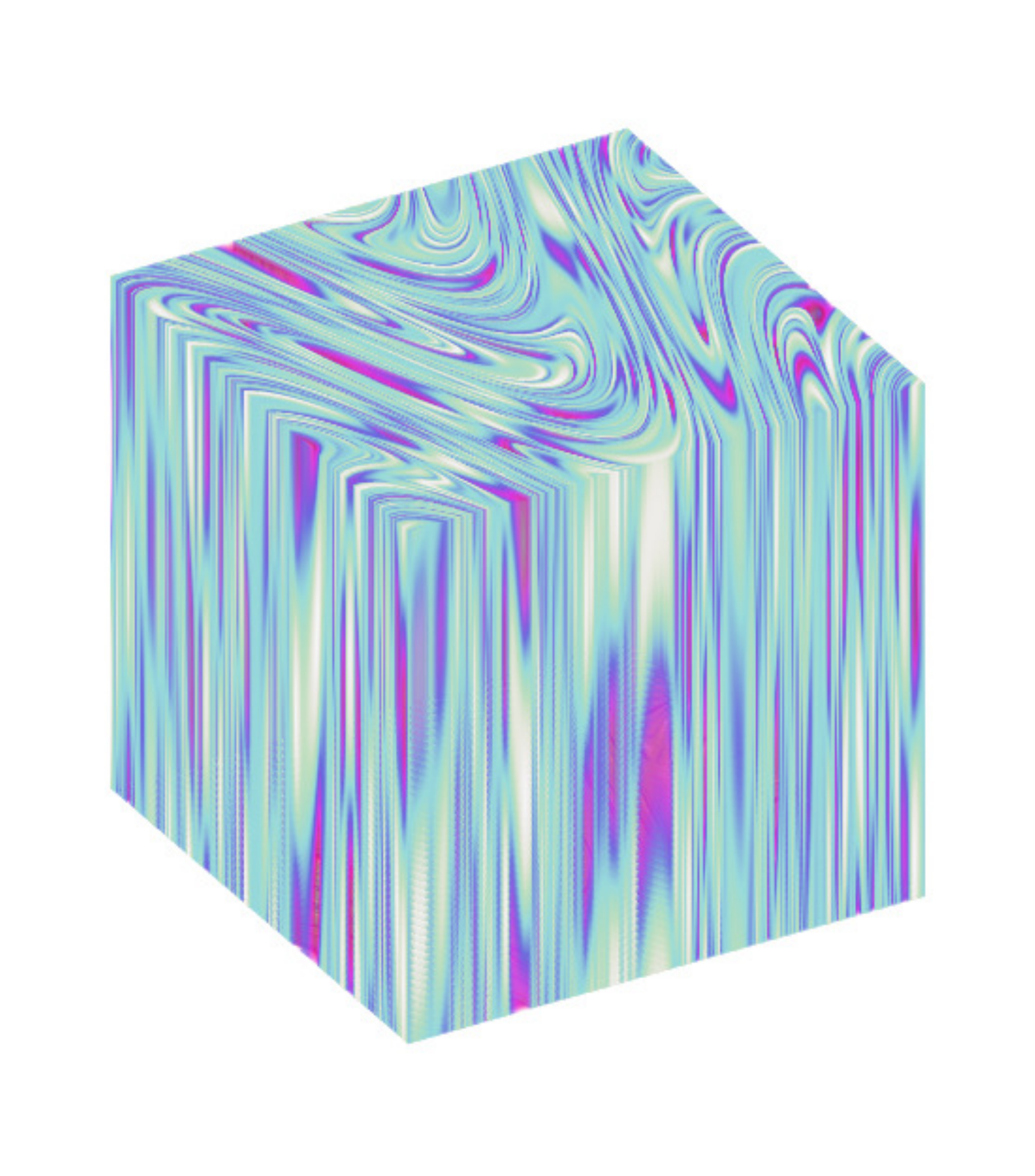}
\caption{(Colour online) Concentration field generated by the map \eqn{2comp}. $n=3$ and
  $\kappa=\ttexp{-4}$. }

\label{fig:2comp}
\end{center}
\end{figure}

The simulations reported in this paper detail the structure of passive
scalars decaying under the combined action of chaotic advection and
(molecular) diffusion in 3D. They confirm that the strange-eigenmode
behaviour \eqn{adle}, which has been well documented in 2D, carries
over to 3D. By contrast with the 2D case, the strange eigenmodes are
found to take three different forms, not only depending on whether the
decay is locally or globally controlled, but also, in the former case,
on the number of positive Lyapunov exponents. Support for this comes
from statistical moments, concentration pdfs and fields, and the
variance decay. 

The simulations confirm that the theoretical predictions of
\citep{hayn-v05} for the decay rate of the concentration
variance are valid in 3D as well as in 2D. In particular, the variance
decay rate for $\kappa \to 0$ in the locally controlled regime is
always given  by $g(0)=-\ell(q_*)$
(Fig.~\ref{fig:vardecay}), the decay rate of the probability that a
line element experience no stretching in the time interval $[0,t]$. As
a result, the variance decays at the same rate for the sine map
\eqn{sine3d} and its inverse \eqn{sine3d}, despite the very different
structures of the scalar fields. The prediction for the variance decay
rate in the globally controlled regime \eqn{localdecay}, which is obtained by applying
homogenisation theory and assuming a large ratio of box size to
characteristic scales, has also been well verified
(Fig.~\ref{fig:vardecay-P}). The transition from local to global
control occurs when this ratio takes the value $P=3$, in agreement
with an estimate obtained by equating the predictions for the decay
rates.

The identical decay rates for the forward and inverse map may seem
surprising. However, this can be explained straightforwardly.
Physically, the variance decay is controlled by a few small fluid
blobs that remain unstretched for long times. The fraction of fluid
occupied by these anomalous regions decreases exponentially at a rate
that is identical for the forward and inverse maps and unrelated to
the number of positive Lyapunov exponents.  Mathematically, the
equality of the decay rates can be established directly, without
resorting to the explicit expression \eqn{localdecay} for the decay
rate. To see this, let us denote by $\mathcal{T}_n$ the action of a
volume-preserving map such as \eqn{sine3d} on the concentration field, $C_n = C(\cdot,n \tau)$.
Immediately after the $n$th advection step 
\beq
C_{n+1}(\bx_{n+1})=\left(\mathcal{T}_n C_{n}\right)(\bx_{n+1})=C_n(\bx_n)
\eeq
while after the $n$th diffusion step (i.e. completion of the full $n$th step)
\beq
C_{n+1}(\bx_{n+1}) = \left(\mathcal{D}  \mathcal{T}_n C_n\right)(\bx_{n+1}),
\eeq
where $\mathcal{D}$ represents the effect of diffusive smoothing,
Using volume preservation, it can be
checked that the adjoint (in $L_2$) of $\mathcal{T}_n$ is
$\mathcal{T}_n^\dagger = \mathcal{T}_n^{-1}$ (i.e.\ $\mathcal{T}_n$ is unitary), while
$\mathcal{D}^\dagger=\mathcal{D}$. Now, after $n+1$ steps of the
forward map, the concentration field is
\beq \lab{TT}
C_{n+1} = \mathcal{D} \mathcal{T}_n \mathcal{D} \mathcal{T}_{n-1} \cdots \mathcal{D} \mathcal{T}_0 C_0,
\eeq
while after $n+1$ steps of the inverse map it is \beq \lab{TT-1}
C_{n+1} = \mathcal{D} \mathcal{T}^{\dagger}_n \mathcal{D}
\mathcal{T}^\dagger_{n-1} \cdots \mathcal{D} \mathcal{T}^\dagger_1
C_0.  \eeq Comparing the operators on the right-hand side of \eqn{TT}
and \eqn{TT-1}, we see that they are almost adjoints of each another,
with differences (relating to the ordering of the $\mathcal{T}_n$ and the
position of $\mathcal{D}$) which are irrelevant to the statistics of
$C_n$ as $n \to \infty$ (since the $\mathcal{T}_n$ are iid). Noting that the decay rate $\lambda$ in
\eqn{adle} can be defined as the largest singular value of these
operators, and that operators which are adjoints of each other have
the same singular values, we conclude that the decay of the forward
and inverse maps are identical.

The prediction \eqn{localdecay} for the decay rate in the locally
controlled regime applies to flows that are sufficiently mixing. The
precise notion of what sufficiently mixing means is not entirely clear,  but we expect it to be close to the
dynamical-system notion of exponential decay of correlations \citep{bala00}. One property
which plays an important role is that of transitivity of the
stretching. This property, which is important for the large-deviation
form \eqn{rate}--\eqn{legendre} of the stretching
statistics \citep{boug-lacr}, requires that the tangent map $A_n$
leave no deterministic directions invariant. As noted earlier, not all
possible 3D extensions of the 2D sine map satisfy this property. To
demonstrate this, we have considered the map
\beq \lab{2comp}
x_{n+1} = x_n + a \sin(y_n + \phi_n), \quad 
y_{n+1} = y_n + b \sin(x_{n+1} + \psi_n), \quad
z_{n+1} = z_n + c \sin(y_{n+1} + \varphi_n),
\eeq
as an alternative to \eqn{sine3d}. The map \eqn{2comp} does not stretch exponentially in the vertical. Consequently the concentration field takes a highly anisotropic form, very different from that obtained for \eqn{sine3d}: the vertical gradients of concentration are much weaker than the horizontal ones, as illustrated by Fig.~\ref{fig:2comp} which displays the concentration from a scalar-decay simulation using \eqn{2comp} and $\kappa=10^{-4}$, and the scalar decay is controlled by horizontal stretching. In this case, the decay rate is obtained from \eqn{localdecay} by considering the horizontal components of \eqn{2comp} only.

\medskip

\noindent
JV acknowledges the support of a Leverhulme Research Fellowship.

\bibliographystyle{apsrev4-1long}
\bibliography{mybib}

\begin{thebibliography}{10}%
\makeatletter
\providecommand \@ifxundefined [1]{%
 \ifx #1\undefined \expandafter \@firstoftwo
 \else \expandafter \@secondoftwo
\fi
}%
\providecommand \@ifnum [1]{%
 \ifnum #1\expandafter \@firstoftwo
 \else \expandafter \@secondoftwo
\fi
}%
\providecommand \enquote [1]{``#1''}%
\providecommand \bibnamefont  [1]{#1}%
\providecommand \bibfnamefont [1]{#1}%
\providecommand \citenamefont [1]{#1}%
\providecommand\href[0]{\@sanitize\@href}%
\providecommand\@href[1]{\endgroup\@@startlink{#1}\endgroup\@@href}%
\providecommand\@@href[1]{#1\@@endlink}%
\providecommand \@sanitize [0]{\begingroup\catcode`\&12\catcode`\#12\relax}%
\@ifxundefined \pdfoutput {\@firstoftwo}{%
 \@ifnum{\z@=\pdfoutput}{\@firstoftwo}{\@secondoftwo}%
}{%
 \providecommand\@@startlink[1]{\leavevmode\special{html:<a href="#1">}}%
 \providecommand\@@endlink[0]{\special{html:</a>}}%
}{%
 \providecommand\@@startlink[1]{%
  \leavevmode
  \pdfstartlink
   attr{/Border[0 0 1 ]/H/I/C[0 1 1]}%
   user{/Subtype/Link/A<</Type/Action/S/URI/URI(#1)>>}%
  \relax
 }%
 \providecommand\@@endlink[0]{\pdfendlink}%
}%
\providecommand \url  [0]{\begingroup\@sanitize \@url }%
\providecommand \@url [1]{\endgroup\@href {#1}{\urlprefix}}%
\providecommand \urlprefix [0]{URL }%
\providecommand \Eprint[0]{\href }%
\@ifxundefined \urlstyle {%
  \providecommand \doi [1]{doi:\discretionary{}{}{}#1}%
}{%
  \providecommand \doi [0]{doi:\discretionary{}{}{}\begingroup
  \urlstyle{rm}\Url }%
}%
\providecommand \doibase [0]{http://dx.doi.org/}%
\providecommand \Doi[1]{\href{\doibase#1}}%
\providecommand \bibAnnote [3]{%
  \BibitemShut{#1}%
  \begin{quotation}\noindent
    \textsc{Key:}\ #2\\\textsc{Annotation:}\ #3%
  \end{quotation}%
}%
\providecommand \bibAnnoteFile [2]{%
  \IfFileExists{#2}{\bibAnnote {#1} {#2} {\input{#2}}}{}%
}%
\providecommand \typeout [0]{\immediate \write \m@ne }%
\providecommand \selectlanguage [0]{\@gobble}%
\providecommand \bibinfo [0]{\@secondoftwo}%
\providecommand \bibfield [0]{\@secondoftwo}%
\providecommand \translation [1]{[#1]}%
\providecommand \BibitemOpen[0]{}%
\providecommand \bibitemStop [0]{}%
\providecommand \bibitemNoStop [0]{.\EOS\space}%
\providecommand \EOS [0]{\spacefactor3000\relax}%
\providecommand \BibitemShut [1]{\csname bibitem#1\endcsname}%
\bibitem{otti89}%
  \BibitemOpen
  \bibfield{author}{%
  \bibinfo {author} {\bibfnamefont{J.~M.}\ \bibnamefont{Ottino}},\ }%
  \emph{\bibinfo {title} {The kinematics of mixing : stretching, chaos, and
  transport}}\ (\bibinfo {publisher} {Cambridge University Press},\ \bibinfo
  {year} {1989})\ p.\ \bibinfo {pages} {364}%
  \bibAnnoteFile{NoStop}{otti89}%
\bibitem{batc59}%
  \BibitemOpen
  \bibfield{author}{%
  \bibinfo {author} {\bibfnamefont{G.~K.}\ \bibnamefont{Batchelor}},\ }%
  \bibfield{title}{%
  \enquote{\bibinfo {title} {Small-scale variation of convected quantities like
  temperature in turbulent fluid. {P}art 1, {G}eneral discussion and the case
  of small convectivity},}\ }%
  \bibfield{journal}{%
  \bibinfo {journal} {J. Fluid Mech.}\ }%
  \textbf{\bibinfo {volume} {5}},\ \bibinfo {pages} {113--133} (\bibinfo {year}
  {1959})%
  \bibAnnoteFile{NoStop}{batc59}%
\bibitem{hayn-angl}%
  \BibitemOpen
  \bibfield{author}{%
  \bibinfo {author} {\bibfnamefont{P.~H.}\ \bibnamefont{Haynes}}\ and\ \bibinfo
  {author} {\bibfnamefont{J.}~\bibnamefont{Anglade}},\ }%
  \bibfield{title}{%
  \enquote{\bibinfo {title} {The vertical-scale cascade in atmospheric tracers
  due to large-scale differential advection},}\ }%
  \bibfield{journal}{%
  \bibinfo {journal} {J. Atmos. Sci.}\ }%
  \textbf{\bibinfo {volume} {54}},\ \bibinfo {pages} {1121--1136} (\bibinfo
  {year} {1997})%
  \bibAnnoteFile{NoStop}{hayn-angl}%
\bibitem{ngan-shep98a}%
  \BibitemOpen
  \bibfield{author}{%
  \bibinfo {author} {\bibfnamefont{K.}~\bibnamefont{Ngan}}\ and\ \bibinfo
  {author} {\bibfnamefont{T.~G.}\ \bibnamefont{Shepherd}},\ }%
  \bibfield{title}{%
  \enquote{\bibinfo {title} {A closer look at chaotic advection in the
  stratosphere. {P}art {I}. {G}eometric structure},}\ }%
  \bibfield{journal}{%
  \bibinfo {journal} {J. Atmos. Sci.}\ }%
  \textbf{\bibinfo {volume} {56}},\ \bibinfo {pages} {4134--4152} (\bibinfo
  {year} {1999})%
  \bibAnnoteFile{NoStop}{ngan-shep98a}%
\bibitem{burg-et-al}%
  \BibitemOpen
  \bibfield{author}{%
  \bibinfo {author} {\bibfnamefont{T.}~\bibnamefont{Burghelea}}, \bibinfo
  {author} {\bibfnamefont{E.}~\bibnamefont{Segre}},\ and\ \bibinfo {author}
  {\bibfnamefont{V.}~\bibnamefont{Steinberg}},\ }%
  \bibfield{title}{%
  \enquote{\bibinfo {title} {Mixing by polymers: experimental test of decay
  regime of mixing},}\ }%
  \bibfield{journal}{%
  \bibinfo {journal} {Phys. Rev. Lett.}\ }%
  \textbf{\bibinfo {volume} {92}},\ \bibinfo {pages} {164501} (\bibinfo {year}
  {2004})%
  \bibAnnoteFile{NoStop}{burg-et-al}%
\bibitem{piko-popo}%
  \BibitemOpen
  \bibfield{author}{%
  \bibinfo {author} {\bibfnamefont{A.}~\bibnamefont{Pikovsky}}\ and\ \bibinfo
  {author} {\bibfnamefont{O.}~\bibnamefont{Popovych}},\ }%
  \bibfield{title}{%
  \enquote{\bibinfo {title} {Persistent patterns in deterministic mixing
  flows},}\ }%
  \bibfield{journal}{%
  \bibinfo {journal} {Europhys. Lett.}\ }%
  \textbf{\bibinfo {volume} {61}},\ \bibinfo {pages} {625--631} (\bibinfo
  {year} {2003})%
  \bibAnnoteFile{NoStop}{piko-popo}%
\bibitem{lebe-turi}%
  \BibitemOpen
  \bibfield{author}{%
  \bibinfo {author} {\bibfnamefont{V.~V.}\ \bibnamefont{Lebedev}}\ and\
  \bibinfo {author} {\bibfnamefont{K.~S.}\ \bibnamefont{Turitsyn}},\ }%
  \bibfield{title}{%
  \enquote{\bibinfo {title} {Passive scalar evolution in peripheral regions},}\
  }%
  \bibfield{journal}{%
  \bibinfo {journal} {Phys. Rev. E}\ }%
  \textbf{\bibinfo {volume} {69}},\ \bibinfo {pages} {036301} (\bibinfo {year}
  {2004})%
  \bibAnnoteFile{NoStop}{lebe-turi}%
\bibitem{salm-hayn}%
  \BibitemOpen
  \bibfield{author}{%
  \bibinfo {author} {\bibfnamefont{H.}~\bibnamefont{Salman}}\ and\ \bibinfo
  {author} {\bibfnamefont{P.~H.}\ \bibnamefont{Haynes}},\ }%
  \bibfield{title}{%
  \enquote{\bibinfo {title} {A numerical study of passive scalar evolution in
  peripheral regions},}\ }%
  \bibfield{journal}{%
  \bibinfo {journal} {Phys. Fluids}\ }%
  \textbf{\bibinfo {volume} {19}},\ \bibinfo {pages} {067101} (\bibinfo {year}
  {2007})%
  \bibAnnoteFile{NoStop}{salm-hayn}%
\bibitem{krai68}%
  \BibitemOpen
  \bibfield{author}{%
  \bibinfo {author} {\bibfnamefont{R.~H.}\ \bibnamefont{Kraichnan}},\ }%
  \bibfield{title}{%
  \enquote{\bibinfo {title} {Small-scale structure of a scalar field convected
  by turbulence},}\ }%
  \bibfield{journal}{%
  \bibinfo {journal} {Phys. Fluids}\ }%
  \textbf{\bibinfo {volume} {11}},\ \bibinfo {pages} {945--953} (\bibinfo
  {year} {1968})%
  \bibAnnoteFile{NoStop}{krai68}%
\bibitem{krai71}%
  \BibitemOpen
  \bibfield{author}{%
  \bibinfo {author} {\bibfnamefont{R.~H.}\ \bibnamefont{Kraichnan}},\ }%
  \bibfield{title}{%
  \enquote{\bibinfo {title} {Inertial-range transfer in two- and
  three-dimensional turbulence},}\ }%
  \bibfield{journal}{%
  \bibinfo {journal} {J. Fluid Mech.}\ }%
  \textbf{\bibinfo {volume} {47}},\ \bibinfo {pages} {525--535} (\bibinfo
  {year} {1971})%
  \bibAnnoteFile{NoStop}{krai71}%
\bibitem{falk-et-al}%
  \BibitemOpen
  \bibfield{author}{%
  \bibinfo {author} {\bibfnamefont{G.}~\bibnamefont{Falkovich}}, \bibinfo
  {author} {\bibfnamefont{K.}~\bibnamefont{Gaw{\c{e}}dzki}},\ and\ \bibinfo
  {author} {\bibfnamefont{M.}~\bibnamefont{Vergassola}},\ }%
  \bibfield{title}{%
  \enquote{\bibinfo {title} {Particles and fields in fluid turbulence},}\ }%
  \bibfield{journal}{%
  \bibinfo {journal} {Rev. Modern Phys.}\ }%
  \textbf{\bibinfo {volume} {73}},\ \bibinfo {pages} {913--975} (\bibinfo
  {year} {2001})%
  \bibAnnoteFile{NoStop}{falk-et-al}%
\bibitem{anto-et-al}%
  \BibitemOpen
  \bibfield{author}{%
  \bibinfo {author} {\bibfnamefont{T.~M.}\ \bibnamefont{Antonsen}}, \bibinfo
  {author} {\bibfnamefont{Z.}~\bibnamefont{Fan}}, \bibinfo {author}
  {\bibfnamefont{E.}~\bibnamefont{Ott}},\ and\ \bibinfo {author}
  {\bibfnamefont{E.}~\bibnamefont{Garcia-Lopez}},\ }%
  \bibfield{title}{%
  \enquote{\bibinfo {title} {The role of chaotic orbits in the determination of
  power spectra of passive scalar},}\ }%
  \bibfield{journal}{%
  \bibinfo {journal} {Phys. Fluids}\ }%
  \textbf{\bibinfo {volume} {8}},\ \bibinfo {pages} {3094--3104} (\bibinfo
  {year} {1996})%
  \bibAnnoteFile{NoStop}{anto-et-al}%
\bibitem{balk-foux}%
  \BibitemOpen
  \bibfield{author}{%
  \bibinfo {author} {\bibfnamefont{E.}~\bibnamefont{Balkovsky}}\ and\ \bibinfo
  {author} {\bibfnamefont{A.}~\bibnamefont{Fouxon}},\ }%
  \bibfield{title}{%
  \enquote{\bibinfo {title} {Universal long-time properties of {L}agrangian
  statistics in the {B}atchelor regime and their application to the passive
  scalar problem},}\ }%
  \bibfield{journal}{%
  \bibinfo {journal} {Phys.\ Rev.\ E}\ }%
  \textbf{\bibinfo {volume} {60}},\ \bibinfo {pages} {4164--4174} (\bibinfo
  {year} {1999})%
  \bibAnnoteFile{NoStop}{balk-foux}%
\bibitem{son99}%
  \BibitemOpen
  \bibfield{author}{%
  \bibinfo {author} {\bibfnamefont{D.~T.}\ \bibnamefont{Son}},\ }%
  \bibfield{title}{%
  \enquote{\bibinfo {title} {Turbulent decay of a passive scalar in the
  {B}atchelor limit: exact results from a quantum mechanical approach},}\ }%
  \bibfield{journal}{%
  \bibinfo {journal} {Phys. Rev. E}\ }%
  \textbf{\bibinfo {volume} {59}},\ \bibinfo {pages} {R3811} (\bibinfo {year}
  {1999})%
  \bibAnnoteFile{NoStop}{son99}%
\bibitem{fere-et-al}%
  \BibitemOpen
  \bibfield{author}{%
  \bibinfo {author} {\bibfnamefont{D.~R.}\ \bibnamefont{Fereday}}, \bibinfo
  {author} {\bibfnamefont{P.~H.}\ \bibnamefont{Haynes}}, \bibinfo {author}
  {\bibfnamefont{A.}~\bibnamefont{Wonhas}},\ and\ \bibinfo {author}
  {\bibfnamefont{J.~C.}\ \bibnamefont{Vassilicos}},\ }%
  \bibfield{title}{%
  \enquote{\bibinfo {title} {Scalar variance decay in chaotic advection and
  {Batchelor}-regime turbulence},}\ }%
  \bibfield{journal}{%
  \bibinfo {journal} {Phys. Rev. E}\ }%
  \textbf{\bibinfo {volume} {65}},\ \bibinfo {pages} {035301} (\bibinfo {year}
  {2002})%
  \bibAnnoteFile{NoStop}{fere-et-al}%
\bibitem{fere-hayn}%
  \BibitemOpen
  \bibfield{author}{%
  \bibinfo {author} {\bibfnamefont{D.~R.}\ \bibnamefont{Fereday}}\ and\
  \bibinfo {author} {\bibfnamefont{P.~H.}\ \bibnamefont{Haynes}},\ }%
  \bibfield{title}{%
  \enquote{\bibinfo {title} {Scalar decay in two-dimensional chaotic advection
  and {B}atchelor-regime turbulence},}\ }%
  \bibfield{journal}{%
  \bibinfo {journal} {Phys. Fluids}\ }%
  \textbf{\bibinfo {volume} {16}},\ \bibinfo {pages} {4359--4370} (\bibinfo
  {year} {2004})%
  \bibAnnoteFile{NoStop}{fere-hayn}%
\bibitem{sche-et-al04}%
  \BibitemOpen
  \bibfield{author}{%
  \bibinfo {author} {\bibfnamefont{A.~A.}\ \bibnamefont{Schekochihin}},
  \bibinfo {author} {\bibfnamefont{P.~H.}\ \bibnamefont{Haynes}},\ and\
  \bibinfo {author} {\bibfnamefont{S.~C.}\ \bibnamefont{Cowley}},\ }%
  \bibfield{title}{%
  \enquote{\bibinfo {title} {Diffusion of passive scalar in a finite-scale
  random flow},}\ }%
  \bibfield{journal}{%
  \bibinfo {journal} {Phys. Rev. E.}\ }%
  \textbf{\bibinfo {volume} {70}},\ \bibinfo {pages} {046304} (\bibinfo {year}
  {2004})%
  \bibAnnoteFile{NoStop}{sche-et-al04}%
\bibitem{tsan-et-al05a}%
  \BibitemOpen
  \bibfield{author}{%
  \bibinfo {author} {\bibfnamefont{Y.-K.}\ \bibnamefont{Tsang}}, \bibinfo
  {author} {\bibfnamefont{T.~M.}\ \bibnamefont{Antonsen}},\ and\ \bibinfo
  {author} {\bibfnamefont{E.}~\bibnamefont{Ott}},\ }%
  \bibfield{title}{%
  \enquote{\bibinfo {title} {Exponential decay of chaotically advected passive
  scalars in the zero diffusivity limit},}\ }%
  \bibfield{journal}{%
  \bibinfo {journal} {Phys. Rev. E}\ }%
  \textbf{\bibinfo {volume} {71}},\ \bibinfo {pages} {066301} (\bibinfo {year}
  {2005})%
  \bibAnnoteFile{NoStop}{tsan-et-al05a}%
\bibitem{hayn-v05}%
  \BibitemOpen
  \bibfield{author}{%
  \bibinfo {author} {\bibfnamefont{P.~H.}\ \bibnamefont{Haynes}}\ and\ \bibinfo
  {author} {\bibfnamefont{J.}~\bibnamefont{Vanneste}},\ }%
  \bibfield{title}{%
  \enquote{\bibinfo {title} {What controls the decay rate of passive scalars in
  smooth random flows?}.}\ }%
  \bibfield{journal}{%
  \bibinfo {journal} {Phys. Fluids}\ }%
  \textbf{\bibinfo {volume} {17}},\ \bibinfo {pages} {097103} (\bibinfo {year}
  {2005})%
  \bibAnnoteFile{NoStop}{hayn-v05}%
\bibitem{mack94}%
  \BibitemOpen
  \bibfield{author}{%
  \bibinfo {author} {\bibfnamefont{R.S.}\ \bibnamefont{Mac{K}ay}},\ }%
  \bibfield{title}{%
  \enquote{\bibinfo {title} {Transport in 3d volume-preserving flows},}\ }%
  \bibfield{journal}{%
  \bibinfo {journal} {J. Nonlin. Sci.}\ }%
  \textbf{\bibinfo {volume} {4}},\ \bibinfo {pages} {329--354} (\bibinfo {year}
  {1994})%
  \bibAnnoteFile{NoStop}{mack94}%
\bibitem{tous-et-al95}%
  \BibitemOpen
  \bibfield{author}{%
  \bibinfo {author} {\bibfnamefont{V.}~\bibnamefont{Toussaint}}, \bibinfo
  {author} {\bibfnamefont{P.}~\bibnamefont{Carri\`ere}},\ and\ \bibinfo
  {author} {\bibfnamefont{F.}~\bibnamefont{Raynal}},\ }%
  \bibfield{title}{%
  \enquote{\bibinfo {title} {A numerical approach to mixing by chaotic
  advection},}\ }%
  \bibfield{journal}{%
  \bibinfo {journal} {Phys. Fluids}\ }%
  \textbf{\bibinfo {volume} {7}},\ \bibinfo {pages} {2834--2844} (\bibinfo
  {year} {1995})%
  \bibAnnoteFile{NoStop}{tous-et-al95}%
\bibitem{tous-et-al00}%
  \BibitemOpen
  \bibfield{author}{%
  \bibinfo {author} {\bibfnamefont{V.}~\bibnamefont{Toussaint}}, \bibinfo
  {author} {\bibfnamefont{P.}~\bibnamefont{Carri\`ere}}, \bibinfo {author}
  {\bibfnamefont{J.}~\bibnamefont{Scott}},\ and\ \bibinfo {author}
  {\bibfnamefont{{J.-N.}}\ \bibnamefont{Gence}},\ }%
  \bibfield{title}{%
  \enquote{\bibinfo {title} {Spectral decay of a passive scalar in chaotic
  mixing},}\ }%
  \bibfield{journal}{%
  \bibinfo {journal} {Phys. Fluids}\ }%
  \textbf{\bibinfo {volume} {12}},\ \bibinfo {pages} {2834--2844} (\bibinfo
  {year} {2000})%
  \bibAnnoteFile{NoStop}{tous-et-al00}%
\bibitem{pier94}%
  \BibitemOpen
  \bibfield{author}{%
  \bibinfo {author} {\bibfnamefont{R.~T.}\ \bibnamefont{Pierrehumbert}},\ }%
  \bibfield{title}{%
  \enquote{\bibinfo {title} {Tracer microstructure in the large-eddy dominated
  regime},}\ }%
  \bibfield{journal}{%
  \bibinfo {journal} {Chaos, Solitons \& Fractals}\ }%
  \textbf{\bibinfo {volume} {4}},\ \bibinfo {pages} {1091--1110} (\bibinfo
  {year} {1994})%
  \bibAnnoteFile{NoStop}{pier94}%
\bibitem{pier00}%
  \BibitemOpen
  \bibfield{author}{%
  \bibinfo {author} {\bibfnamefont{R.~T.}\ \bibnamefont{Pierrehumbert}},\ }%
  \bibfield{title}{%
  \enquote{\bibinfo {title} {Lattice models of advection-diffusion},}\ }%
  \bibfield{journal}{%
  \bibinfo {journal} {Chaos}\ }%
  \textbf{\bibinfo {volume} {10}},\ \bibinfo {pages} {61--74} (\bibinfo {year}
  {2000})%
  \bibAnnoteFile{NoStop}{pier00}%
\bibitem{hall-yuan}%
  \BibitemOpen
  \bibfield{author}{%
  \bibinfo {author} {\bibfnamefont{G.}~\bibnamefont{Haller}}\ and\ \bibinfo
  {author} {\bibfnamefont{G.}~\bibnamefont{Yuan}},\ }%
  \bibfield{title}{%
  \enquote{\bibinfo {title} {Lagrangian coherent structures and mixing in
  two-dimensional turbulence},}\ }%
  \bibfield{journal}{%
  \bibinfo {journal} {Physica D}\ }%
  \textbf{\bibinfo {volume} {147}},\ \bibinfo {pages} {352--370} (\bibinfo
  {year} {2000})%
  \bibAnnoteFile{NoStop}{hall-yuan}%
\bibitem{fann-et-al04}%
  \BibitemOpen
  \bibfield{author}{%
  \bibinfo {author} {\bibfnamefont{A.}~\bibnamefont{Fannjiang}}, \bibinfo
  {author} {\bibfnamefont{S.}~\bibnamefont{Nonnenmacher}},\ and\ \bibinfo
  {author} {\bibnamefont{L.Wolowski}},\ }%
  \bibfield{title}{%
  \enquote{\bibinfo {title} {Dissipation time and decay of correlations},}\ }%
  \bibfield{journal}{%
  \bibinfo {journal} {Nonlinearity}\ }%
  \textbf{\bibinfo {volume} {17}},\ \bibinfo {pages} {1481--1508} (\bibinfo
  {year} {2004})%
  \bibAnnoteFile{NoStop}{fann-et-al04}%
\bibitem{ott02}%
  \BibitemOpen
  \bibfield{author}{%
  \bibinfo {author} {\bibfnamefont{E.}~\bibnamefont{Ott}},\ }%
  \emph{\bibinfo {title} {Chaos in dynamical systems}},\ \bibinfo {edition}
  {2nd}\ ed.\ (\bibinfo {publisher} {Cambridge University Press},\ \bibinfo
  {year} {2002})\ p.\ \bibinfo {pages} {478}%
  \bibAnnoteFile{NoStop}{ott02}%
\bibitem{v10}%
  \BibitemOpen
  \bibfield{author}{%
  \bibinfo {author} {\bibfnamefont{J.}~\bibnamefont{Vanneste}},\ }%
  \bibfield{title}{%
  \enquote{\bibinfo {title} {Estimating generalized {L}yapunov exponents for
  products of random matrices},}\ }%
  \bibfield{journal}{%
  \bibinfo {journal} {Phys. Rev. E}\ }%
  \textbf{\bibinfo {volume} {81}},\ \bibinfo {pages} {036701} (\bibinfo {year}
  {2010})%
  \bibAnnoteFile{NoStop}{v10}%
\bibitem{sukh-pier02a}%
  \BibitemOpen
  \bibfield{author}{%
  \bibinfo {author} {\bibfnamefont{J.}~\bibnamefont{Sukhatme}}\ and\ \bibinfo
  {author} {\bibfnamefont{R.~T.}\ \bibnamefont{Pierrehumbert}},\ }%
  \bibfield{title}{%
  \enquote{\bibinfo {title} {Decay of passive scalars under the action of
  single scale smooth velocity fields in bounded two-dimensional domains: from
  non-self-similar probability distribution functions to self-similar
  eigenmodes},}\ }%
  \bibfield{journal}{%
  \bibinfo {journal} {Phys. Rev. E}\ }%
  \textbf{\bibinfo {volume} {66}},\ \bibinfo {pages} {056302} (\bibinfo {year}
  {2002})%
  \bibAnnoteFile{NoStop}{sukh-pier02a}%
\bibitem{thif-chil}%
  \BibitemOpen
  \bibfield{author}{%
  \bibinfo {author} {\bibfnamefont{{J.-L.}}\ \bibnamefont{Thiffeault}}\ and\
  \bibinfo {author} {\bibfnamefont{S.}~\bibnamefont{Childress}},\ }%
  \bibfield{title}{%
  \enquote{\bibinfo {title} {Chaotic mixing in a torus map},}\ }%
  \bibfield{journal}{%
  \bibinfo {journal} {Chaos}\ }%
  \textbf{\bibinfo {volume} {13}},\ \bibinfo {pages} {502--507} (\bibinfo
  {year} {2003})%
  \bibAnnoteFile{NoStop}{thif-chil}%
\bibitem{Note1}%
  \BibitemOpen
  \bibinfo {note} {``Kraichan--Kazantsev flows'' are obtained in the limit of
  zero correlation time for the velocity field (i.e.\ a white-in-time velocity
  field); ``renewing flows'' or ``renovating flows'' are described by
  independent, identically distributed random processes that become completely
  decorrelated after a finite time interval.}%
  \bibAnnoteFile{Stop}{Note1}%
\bibitem{majd-kram}%
  \BibitemOpen
  \bibfield{author}{%
  \bibinfo {author} {\bibfnamefont{A.~J.}\ \bibnamefont{Majda}}\ and\ \bibinfo
  {author} {\bibfnamefont{P.~R.}\ \bibnamefont{Kramer}},\ }%
  \bibfield{title}{%
  \enquote{\bibinfo {title} {Simplified models for turbulent diffusion: theory,
  numerical modelling and physical phenomena},}\ }%
  \bibfield{journal}{%
  \bibinfo {journal} {Phys. Rep.}\ }%
  \textbf{\bibinfo {volume} {314}},\ \bibinfo {pages} {237--574} (\bibinfo
  {year} {1999})%
  \bibAnnoteFile{NoStop}{majd-kram}%
\bibitem{v06a}%
  \BibitemOpen
  \bibfield{author}{%
  \bibinfo {author} {\bibfnamefont{J.}~\bibnamefont{Vanneste}},\ }%
  \bibfield{title}{%
  \enquote{\bibinfo {title} {Intermittency of passive-scalar decay: strange
  eigenmodes in random shear flows},}\ }%
  \bibfield{journal}{%
  \bibinfo {journal} {Phys. Fluids}\ }%
  \textbf{\bibinfo {volume} {18}},\ \bibinfo {pages} {087108} (\bibinfo {year}
  {2006})%
  \bibAnnoteFile{NoStop}{v06a}%
\bibitem{Note2}%
  \BibitemOpen
  \bibinfo {note} {Note that \protect \citep {balk-foux} predicts a decay rate
  for $d=3$ which differs from $\gamma \sim - \ell (q_*) = g(0)$ and depends on
  whether a flow has one or two expanding directions.}%
  \bibAnnoteFile{Stop}{Note2}%
\bibitem{bala00}%
  \BibitemOpen
  \bibfield{author}{%
  \bibinfo {author} {\bibfnamefont{V.}~\bibnamefont{Baladi}},\ }%
  \emph{\bibinfo {title} {Positive transfer operators and decay of
  correlations}},\ Vol.~\bibinfo {volume} {16}\ (\bibinfo {publisher} {World
  Scientific},\ \bibinfo {year} {2000})%
  \bibAnnoteFile{NoStop}{bala00}%
\bibitem{boug-lacr}%
  \BibitemOpen
  \bibfield{author}{%
  \bibinfo {author} {\bibfnamefont{P.}~\bibnamefont{Bougerol}}\ and\ \bibinfo
  {author} {\bibfnamefont{J.}~\bibnamefont{Lacroix}},\ }%
  \emph{\bibinfo {title} {Products of random matrices with applications to
  Schr\"odinger operators}}\ (\bibinfo {publisher} {Birkh\"auser},\ \bibinfo
  {year} {1985})\ p.\ \bibinfo {pages} {283pp}%
  \bibAnnoteFile{NoStop}{boug-lacr}%
\end{thebibliography}%

\end{document}